\documentstyle[amsmath,amssymb,graphicx]{article}

\def\be{\begin{eqnarray}}
\def\ee{\end{eqnarray}}
\def\nn{\nonumber}

\def\A{{\cal A}}
\def\B{{\cal B}}
\def\I{{\cal I}}
\def\J{{\cal J}}

\def\[{\phantom.[}
\def\underd{\boxed }

\hoffset= - 0.95in         % switch off for draft style

\begin{document}

\hfill ITEP/TH-13/16

\hfill IITP/TH-10/16

\bigskip

\bigskip

\centerline{\Large{Factorization of differential expansion
for antiparallel double-braid knots}}

%\centerline{\Large{
%and exclusive Racah matrices for rectangular representations
%}}

\bigskip

\centerline{\bf  A.Morozov }

\bigskip

{\footnotesize
\centerline{{\it
ITEP, Moscow 117218, Russia}}

\centerline{{\it
Institute for Information Transmission Problems, Moscow 127994, Russia
}}

\centerline{{\it
National Research Nuclear University MEPhI, Moscow 115409, Russia
}}
}

\bigskip

\bigskip

\centerline{ABSTRACT}

\bigskip

{\footnotesize
Continuing the quest for exclusive Racah matrices, which are needed for
evaluation of colored arborescent-knot polynomials in Chern-Simons theory,
we suggest to extract them from a new kind of a double-evolution --
that of the antiparallel double-braids, which is a simple two-parametric
family of two-bridge knots, generalizing the one-parametric family of
twist knots.
In the case of rectangular representations $R=[r^s]$ we found an evidence
that the corresponding differential expansion miraculously factorizes
and can be obtained from that for the twist knots.
This reduces the problem of rectangular exclusive Racah to constructing
the answers for just a few twist knots.
We develop a recent conjecture on the structure of differential expansion
for the simplest members of this family (the trefoil and the figure-eight knot)
and provide the exhaustive answer
for the first unknown case of $R=[33]$.
The answer includes HOMFLY of arbitrary twist and double-braid knots
and Racah matrices $\bar S$ and $S$ -- what allows to calculate
$[33]$-colored polynomials
for arbitrary arborescent (double-fat) knots.
For generic rectangular representations described in detail are only
the contributions of the single- and two-floor pyramids,
the way to proceed is explicitly illustrated by the examples of $R=[44]$
and $R=[55]$.
This solves the difficult part of the problem, but
the last tedious step towards explicit formulas for generic
exclusive rectangular Racah matrices still remains to be made.

}

\bigskip

\section{Introduction}

Construction of knot polynomials \cite{knotpols,Con} is currently
at the front-line of  theoretical physics,
because this is an {\it exactly solvable} problem in quantum field theory,
which remains {\it unsolved} for years, despite tremendous effort by many distinguished researchers.
Far ago it was reformulated in terms of Chern-Simons theory \cite{CS}-\cite{GSV},
where (at least for the knots in $S^3$) it is basically reduced to a free-field calculation,
and a formal answer is provided \cite{RT}-\cite{MoSm} in terms of representation theory of quantum groups.
This, however, does not help to obtain {\it explicit answers}, except for the simplest situations,
covered by the well-known databases \cite{katlas}.
The problem is that group-theory methods themselves are undeveloped, moreover, the crucial
quantities, starting from Racah matrices, are ambiguously defined in most interesting cases
and thus do not attract attention of pure mathematicians.
The ambiguities drop out of the final answers for knot polynomials, but intermediate steps
involve less-invariant objects -- what can in fact be interpreted as a kind of a new gauge
invariance of some effective field theory \cite{MMfam},
arising on the way from the fundamental Chern-Simons to Wilson-loop observables.
This somewhat unexpected twist of the story makes it even more similar to "real" Yang-Mills
theories, like QCD, and confirms expectations that the study of exactly-solvable models
can shed light on the more complicated physical problems.

Recent revival of interest and new fast progress in $2d$ conformal theories (CFT) \cite{CFT}-\cite{DIM}
attracted new attention to $3d$ Chern-Simons, which is its closest relative --
and naturally caused a progress there, which is, however, not so spectacular yet.
The difference is that knot polynomials are exact {\it non-perturbative} quantities,
associated with modular transformations of non-perturbative conformal blocks,
which are still far from being well-studied in conformal theory as well.
Fortunately, modular transformations are {\it simpler} than conformal blocks themselves,
and one can proceed with knot polynomials even when conformal-block issues remain unsolved.
Recent achievements in this direction are largely based on the new version \cite{MMMkn12}-\cite{Garou}
of the Reshetikhin-Turaev (RT) formalism, when quantum ${\cal R}$-matrices
in the space of representations, rather than in representation spaces,
and Clebsh-Gordan coefficients are substituted by Racah and matrices
(and their more complicated convolutions, known as {\it mixing} matrices).
An early example of the strength of such approach was provided by the celebrated
Rosso-Jones formula \cite{RJ}-\cite{GGS}, which fully solves the problem of colored knot polynomials
for a distinguished case of torus knots.
However, despite a considerable progress, made in above references,
nothing comparably impressive is yet achieved beyond torus knots --
the problem turns to be extremely complicated.
A new hope appeared with the introduction of the special class of {\it double-fat} knots
in \cite{mmmrv}, where knot polynomials are presumably made from monodromy matrices
of 4-point conformal blocks -- and thus the CFT methods can be directly applied.
This class is rather rich, in includes all the two-bridge and pretzel knots,
moreover, it appeared to coincide with the {\it arborescent knots}, well known
in mathematical literature \cite{arb}.
In general such knots are made by contractions of "fingers" and "propagators",
with just four lines/strands inside, and their knot polynomials can be considered
as correlators in some new effective field theory \cite{MMfam}, which
is in fact a gauge theory, as we already mentioned.
What is needed for arborescent knots is just a pair of Racah matrices,
\be
S: \ \ \ \ \Big((\bar R\otimes  R) \otimes R\ \longrightarrow\  R\Big)
\ \ \longrightarrow \ \ \Big(\bar R\otimes (R\otimes R)\ \longrightarrow \ R\Big)
\nn \\
\bar S: \ \ \ \ \Big((R\otimes\bar R) \otimes R\ \longrightarrow\  R\Big)
\ \ \longrightarrow \ \ \Big(R\otimes (\bar R\otimes R)\ \longrightarrow \ R\Big)
\ee
called {\it exclusive}, to distinguish them from {\it exclusive} ones, where the
"final state" is arbitrary representation $Q\in R\otimes R\otimes \bar R$, not just $R$.
The problem, however, is that they are needed in arbitrary representation $R$,
if one wants to calculate $R$-colored knot polynomials.
Modern group theory is incapable to provide the answers beyond pure
symmetric and antisymmetric representations \cite{sympols} --
the record results, obtained
by various direct methods are for $R=[2,1]$ \cite{GJ}
and $R=[2,2]$ \cite{mmms3} (see also \cite{mmms1} and \cite{mmms2} for some
{\it inclusive} Racah matrices for $R=[3,1]$ and $R=[2,2]$ respectively).
Further progress on these lines seems to be beyond the current computer capacities.

However, the answers, when known, are pretty simple, and it is clear that they can be
found/guessed by some other, indirect methods.
One of the obvious  ideas is to extract the Racah matrices from the answers for
some relatively simple knots and then apply them to calculations of generic (at least,
arborescent) knots.
An illustration of this idea was already given in \cite{mmms3} with the example
of representation $R=[22]$, where   $S$ and $\bar S$ were extracted from
{\it exclusive} 3-strand Racah matrices,
previously found from the difficult direct calculation in \cite{mmms2},
using the two-parametric family, which is simultaneously arborescent and 3-strand.
Despite impressive, this example just expands the small piece of knowledge about
a given $R$ into a much bigger knowledge -- but only about the same $R$.
Given problems with higher $R$, this is not enough.

An example of a very different kind was provided in \cite{rect41}.
One of the recent discoveries about knot polynomials is their spectacular
internal structure, known as {\it differential expansion} (DE) \cite{IMMMfe}-\cite{Konodef}.
It goes back to discovery of "differentials" in \cite{DGR},
based on the understanding of Khovanov-Rozansky calculus \cite{Kho} --
a conceptually important alternative to RT approach.
DE somehow lies at the intersection of two different formalisms
and, non-surprisingly, is a very powerful idea -- unfortunately, underestimated and undeveloped.
Still, whenever applied, it proves effective.
In fact, the matrices $S$ and $\bar S$ for all (anti)symmetric representations
in  \cite{sympols} were obtained as generalization of the answers for the figure-eight
$4_1$ and other twist knots in \cite{IMMMfe} and \cite{MMMevo} -- which were originally
{\it guessed} from the study of what later became the differential expansion.
The suggestion of \cite{rect41} was to repeat this trick on the new level of knowledge:
to look at newly available results, reformulate them in terms of differential
expansion and then, hopefully, generalize -- producing absolutely new knowledge,
or at least conjectures, about the unknown.
In this respect \cite{rect41} was a success:
the Rosso-Jones answers for the trefoil for {\it all rectangular} representations $R$
were reformulated in terms of differential expansion, what allowed to conjecture
the rectangularly-colored HOMFLY for figure eight knot and, most important,
to conjecture the general shape of differential expansion in this case for
all defect-zero \cite{Konodef} knots.
The purpose of this paper is to further built on this success
and make a new step towards finding the matrices $\bar S$ and $S$ for arbitrary
rectangular representations.
In fact just one more guess remains to be done on this way after the present paper --
but first an independent examination of the already made conjectures is highly
desirable.

\bigskip

Since \cite{arthdiff} it is known, that differential expansion is much simpler for
antiparallel than antiparallel braids, i.e. for twist rather than torus knots (even 2-strand).
The latter have non-vanishing defect \cite{Konodef} and instead possess non-trivial next levels
of DE \cite{arthdiff}, i.e. an additional structure, which still needs to be understood.
Here we concentrate on the simpler -- antiparallel -- story, where defect is always zero and
the structure of DE is fully described in \cite{rect41}.
We demonstrate that it exhibits a new factorization property,
relating expansion coefficients for multi-braid knots through those for twist knots.
This opens a way to explicitly describe a double evolution, extract exclusive Racah matrices
$S$ and $\bar S$ by a variation of the idea in \cite{mmms2} and apply the technique of
\cite{mmmrv}-\cite{mmms2} to find HOMFLY polynomials for arbitrary arborescent knots.
in arbitrary rectangular representation $R=r^s]$.
The success of this program -- once factorization is discovered --
depends on the possibility to guess the general shape of the DE {\it coefficients}
for twist knots,
which hopefully will be possible in the near future.

Generalization to non-rectangular diagrams faces additional (perhaps, related) problems:
the structure of DE is more sophisticated, even for trefoil and the figure eight,
and non-trivial multiplicities arise, turning Racah matrices in more sophisticated
operators, which do not possess any canonical matrix form.
Moreover, additional care is needed \cite{MMfam} in this case to formulate the arborescent calculus
of \cite{mmmrv} for multi-finger knots -- the interaction vertices in the corresponding
effective "field theory" are also not canonically defined (or "non-local").
Thus non-rectangular case will be further elaborated on elsewhere.

\bigskip

Despite there is hardly any need to advertise the long-standing problem of
calculating at least some non-symmetrically-colored HOMFLY for at least some knots --
any result here is a breakthrough -- we begin in s.\ref{devo}
from explaining that the particular two-parametric family
of antiparallel double-braids is especially important  at the present stage of knowledge:
if found, it provides exclusive Racah matrices and thus allows to calculate for
arbitrary arborescent knots, what is already quite something.
The idea is further illustrated by the example of the fundamental representation
in s.\ref{funddevo}.
We switch to DE only in s.\ref{detwist22}, where the coefficients of the expansion,
structured in \cite{rect41}, are found for the twist knots, by
applying the evolution method \cite{MMMevo} to the known $[22]$-colored-HOMFLY from \cite{mmms2}.
The culmination comes in s.\ref{factdb22}, where with available examples
we discover and illustrate the factorization property of DE for antiparallel double braids --
and reproduce from the approach of s.\ref{devo} the exclusive Racah matrices
$\bar S_{22}$, recently found from a very different double evolution in \cite{mmms3}.
%In s.\ref{factmb22} we further elaborate on factorization -- this time for generic
%antiparallel multi-braids -- and
In s.\ref{plan} we make a formal conjecture, that this property holds for arbitrary
rectangular representations.
%As explained in s.\ref{plan},
Given all the recent achievements after \cite{mmmrv} this basically
reduces rectangular arborescent calculus to finding the DE coefficients in \cite{rect41}
for the rather simple one-parametric family of twist knots.
For them a separate conjecture is needed,
of which we cook up just a prototype in s.\ref{Ftwist}:
we make it in full generality for the single-floor pyramids.
At two floors we consider only the first unknown example of $R=[33]$.
Further generalizations look rather straightforward, but  are left
for the future work.

%?????For them we cook up a separate conjecture in s.\ref{Ftwist},
%which is the second culmination in this text.
%????

\bigskip

Throughout the text we use the standard notation $\{x\}=x-x^{-1}$ and $[n]=\frac{\{q^n\}}{\{q\}}$.
Question signs over equalities denote {\it conjectures}.

%???

\section{Exclusive Racah matrices from double evolutions
\label{devo}}

In \cite{mmms3} exclusive Racah matrices $S$ were extracted from the double evolution family of
3-strand knots: they diagonalize the double evolution matrix.
 Then $\bar S$ is obtained from
 \be
 \bar S = \bar T^{-1} S T^{-1} S^\dagger \bar T^{-1}
 \label{SvsbS}
 \ee
with diagonal matrices $T$ and $\bar T$, made from the eigenvalues of the relevant ${\cal R}$-matrices.
Instead, the double evolution family of double-braid two-bridge knots defines $\bar S$ directly:
in the notation of  \cite{mmmrv}
\be
H^{\{m,n\}}_R = d_R\cdot \Big(\bar S \bar T^{2m} \bar S \bar T^{2n} \bar S)_{\emptyset\emptyset}
= \sum_{\bar X,\bar Y\in R\otimes\bar R}
\frac{\sqrt{d_{\bar X}d_{\bar Y}}}{d_R} \lambda_{\bar X}^{2m}\lambda_{\bar Y}^{2n} \bar S_{\bar X\bar Y}
\label{debarS}
\ee
and $S$ is then extracted from (\ref{SvsbS}) as a diagonalizing matrix of
$\bar T \bar S \bar T$.

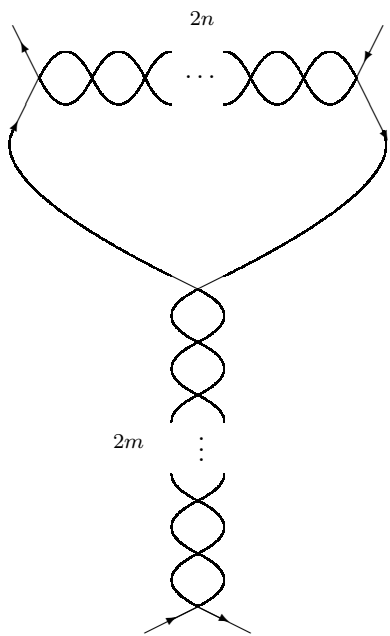
\begin{picture}(200,270)(-230,-230)
\qbezier(-40,0)(-50,20)(-60,0)
\qbezier(-40,0)(-50,-20)(-60,0)
\qbezier(-20,0)(-30,20)(-40,0)
\qbezier(-20,0)(-30,-20)(-40,0)
\qbezier(-20,0)(-15,10)(-10,10)
\qbezier(-20,0)(-15,-10)(-10,-10)
\put(-5,0){\mbox{$\ldots$}}
\qbezier(10,10)(15,10)(20,0)
\qbezier(10,-10)(15,-10)(20,0)
\qbezier(20,0)(30,20)(40,0)
\qbezier(20,0)(30,-20)(40,0)
\qbezier(40,0)(50,20)(60,0)
\qbezier(40,0)(50,-20)(60,0)
\put(-60,0){\line(-1,2){10}}
\put(-60,0){\line(-1,-2){10}}
\put(60,0){\line(1,2){10}}
\put(60,0){\line(1,-2){10}}
\qbezier(0,-80)(-20,-90)(0,-100)
\qbezier(0,-80)(20,-90)(0,-100)
\qbezier(0,-100)(-20,-110)(0,-120)
\qbezier(0,-100)(20,-110)(0,-120)
\qbezier(0,-120)(-10,-125)(-10,-130)
\qbezier(0,-120)(10,-125)(10,-130)
\put(0,-145){\mbox{$\vdots$}}
\qbezier(0,-160)(-10,-155)(-10,-150)
\qbezier(0,-160)(10,-155)(10,-150)
\qbezier(0,-160)(-20,-170)(0,-180)
\qbezier(0,-160)(20,-170)(0,-180)
\qbezier(0,-180)(-20,-190)(0,-200)
\qbezier(0,-180)(20,-190)(0,-200)
\put(0,-80){\line(-2,1){10}}
\put(0,-80){\line(2,1){10}}
\put(0,-200){\line(-2,-1){10}}
\put(0,-200){\line(2,-1){10}}
\put(0,-200){\line(-2,-1){20}}
\put(0,-200){\line(2,-1){20}}
\qbezier(-10,-75)(-80,-40)(-70,-20)
\qbezier(10,-75)(80,-40)(70,-20)
\put(-10,-205){\vector(2,1){2}}
\put(10,-205){\vector(2,-1){2}}
\put(-65,10){\vector(-1,2){2}}
\put(65,10){\vector(-1,-2){2}}
\put(-70,-20){\vector(1,2){2}}
\put(70,-20){\vector(1,-2){2}}
\put(-3,20){\mbox{\footnotesize$2n$}}
\put(-32,-140){\mbox{\footnotesize $2m$}}
%%
%\put(-11,20){\mbox{\footnotesize$2k+1$}}
%%
%\put(-30,-140){\mbox{\footnotesize $n$}}
%%
%\put(-2,-230){\mbox{I}}
%
\end{picture}

Thus a knowledge of rectangular HOMFLY for the double-braid family can  be used
to obtain these Racah matrices in rectangular representations.
In particular, this provides an alternative derivation for $R=[22]$ --
and coincidence with the result of  \cite{mmms3} can serve as a check of
our factorization hypotheses about the double-braid.

\section{Fundamental representation
\label{funddevo}}

For double braids  of above type (two antiparallel braids with even crossing numbers
the normalized fundamental HOMFLY are equal to
\be
H^{\{m,n\}}_{[1]} = 1+ G_{[1]}^{(m,n)}\{Aq\}\{A/q\}
=  1 + \frac{(A^{2m}-1)(A^{2n}-1)}{(A^2-1)(A^{-2}-1)}\,\{Aq\}\{A/q\}
\label{fundmn}
\ee
with $A$-independent factor $F_{[1]}^{(m,n)}$.
Thus Alexander polynomial $Al_{[1]}^{(m,n)} = 1+mn\{q\}^2$ has degree one,
and {\it defect} \cite{Konodef} of the differential expansion is zero
for the entire family.
This means that we can use the conjecture of \cite{rect41}
for the shape of differential expansions for rectangular representations
of the defect-zero knots.
What needs to be found are the $A,q$-dependent coefficients ($F$-factors),
which should be guessed from the limited knowledge of HOMFLY for particular
members of the family:
\be
\begin{array}{c|cccccccccccccccc}
2m & \ldots &   & -8 & -6 & -4 & -2 & 0 & 2 & 4 & 6 & 8 &   & \ldots \\
2n &&&&&&&&&&&&& \\ \hline
\ldots&&&&&&&&&&&&&&&& \\
%10 &&&&&&&U&&&&&&&& \\
8  &&&&&&10_1&U&9_2&&&&&&& \\
6  &&&&&10_3&8_1&U&7_2&9_5&&&&&& \\
4  &&&&10_3&8_3&6_1&U&5_2&7_4&9_5&&&&& \\
2  &&&10_1&8_1&6_1&4_1&U&3_1&5_2&7_2&9_2&&&& \\
0   && &U&U&U&U&U&U&U&U&U&  \\
-2   &&&9_2&7_2&5_2&3_1&U&4_1&6_1&8_1&10_1&&&& \\
-4   &&&&9_5&7_4&5_2&U&6_1&8_3&10_3&&&&& \\
-6   &&&&&9_5&7_2&U&8_1&10_3&&&&&& \\
-8   &&&&&&9_2&U&10_1&&&&&&& \\
%-10  &&&&&&&U&&&&&&&& \\
\ldots &
\end{array}
\label{dbraidtable}
\ee
$U$ means the unknot (all $F$-factors are zero),
the lines/columns with $2m$ or $2n=\pm 2$ contain twist knots.
The table has obvious symmetries $m\leftrightarrow n$ and $(m,n)\leftrightarrow (-m,-n)$,
%(in the latter case the knot turns into its mirror, and the change $A,q\longrightarrow A^{-1},q^{-1}$
%should be made in knot polynomial),
thus there are not too many different knots in the table.
Fortunately, some colored HOMFLY are also available beyond 10 crossings
due to powerful family method of \cite{MMfam}.

\bigskip

For the fundamental representation $R=[1]$ with $d_{[1]}=[N]$ we get from (\ref{fundmn})
\be
 \frac{\sqrt{d_Xd_Y}}{d_R}\, \bar S_{XY} = \left(\begin{array}{cc} 1 & 0 \\ 0 & 0 \end{array}\right)
- \frac{\{Aq\}\{A/q\}}{\{A\}^2}\left(\begin{array}{cc} 1 & -1  \\ -1  & 1 \end{array}\right)
= \frac{1}{[N]^2} \left(\begin{array}{cc} 1 & \ \ [N-1][N+1] \\ \[N-1][N+1] & -[N-1][N+1]\end{array}\right)
\ee
i.e.
\vspace{-0.2cm}
\be
\bar S_{[1]} = \frac{1}{[N]}
\left(\begin{array}{cc} 1 & \sqrt{[N-1][N+1]} \\ \\ \sqrt{[N-1][N+1]} & -1 \end{array}\right)
\ee

\section{Twist knots for $R=[22]$
\label{detwist22}}

According to \cite{rect41}, for defect-zero knots ${\cal K}^{(0)}$,
\be
H_{[22]}^{{\cal K}^{(0)}} = 1 + [2]^2  Z_{[22]}^{(0)} \cdot F_{[1]}^{{\cal K}^{(0)}}\!(A)
+ [3]Z_{[22]}^{(0)}\Big(Z_{[22]}^{(1)}\cdot F_{[2]}^{{\cal K}^{(0)}}\!(A,q)
+ Z_{[22]}^{(-1)}\cdot F_{[2]}^{{\cal K}^{(0)}}\!(A,q^{-1}) \Big) + \nn \\
+ Z_{[22]}^{(1)}Z_{[22]}^{(0)}Z_{[22]}^{(-1)}\Big([2]^2 {\cal F}_3^{{\cal K}^{(0)}}\!(A,q)
+Z_{[22]}^{(0)}\cdot {\cal F}_4^{{\cal K}^{(0)}}\!(A,q)\Big)
\ee
The two $F$-factors in the first line  are known from differential expansions for
symmetric representations, the two ${\cal F}$-factors in the second line are some
$q$-deformations of the cube and the forth power of $F_{[1]}(A)$ -- both symmetric
under the change $q\leftrightarrow q^{-1}$ (because of the transposition symmetry of $R=[22]$),
but very different from, say, $F_{[3]}(A,q=1)$ and $F_{[4]}(A,q=1)$.
They should, however, satisfy the evolution  rule
with six eigenvalues $\bar T_{[22]} = {\rm diag}\Big(1,\ -A,\ q^{ 2}A^2,\ q^{-2}A^2,\ -A^3,\ A^4\Big)$.
%\vspace{-0.2cm}
\be
\ldots \nn  \\
{\cal F}_3^{[4_1]} = 1 \nn \\
{\cal F}_3^{[U]} = 0 \nn \\
{\cal F}_3^{[3_1]} = -A^6 \nn \\
{\cal F}_3^{[5_2]} = -A^6\left(A^6 + \frac{[6]}{[2]}A^4 + [3]A^2+1\right) \nn \\
{\cal F}_3^{[7_2]} = -A^6\left(A^{12} + \frac{[6]}{[2]}A^{10}+ \frac{[8][6]}{[4][2]}A^8
+ [7]A^6 + \frac{[4][3]}{[2]}A^4+ [3]A^2+1\right) \nn \\
{\cal F}_3^{[9_2]} = -A^6\left(A^{18} + \frac{[6]}{[2]}A^{16}+\frac{[8][6]}{[4][2]}A^{14}
+\frac{[10][8]}{[4][2]}A^{12} +\frac{[10][6]}{[5]}A^{10} + \frac{[8][3]}{[2]}A^8
+ \underbrace{\frac{[5][4]}{[2]}A^6 + \frac{[4][3]}{[2]}A^4+ [3]A^2+1}_{
\frac{1}{(1-q^2A^2)(1-A^2)(1-q^{-2}A^2)}
}\right) \nn \\
\ldots \nn \\
\nn
\ee
\be
\ldots \nn \\ \nn \\
{\cal F}_4^{[4_1]} = 1 \nn \\
{\cal F}_4^{[U]} = 0 \nn \\
{\cal F}_4^{[3_1]} = A^8 \nn \\
{\cal F}_4^{[5_2]} = A^8\left(A^8 + [2]^2A^6+ \frac{[4][3]}{[2]}A^4 + [2]^2A^2+1\right) \nn \\
{\cal F}_4^{[7_2]} = A^8\left(A^{16} + [2]^2A^{14}+ ([3]^2+1)A^{12} + [4]^2A^{10}
+ \Big(\frac{[6][5]}{[2]}+[3]+1\Big)A^8 + [4]^2A^6 + ([3]^2+1)A^4 + [2]^2A^2+1\right) \nn \\
{\cal F}_4^{[9_2]} = A^8\left(A^{24}  + [2]^2A^{22} + ([3]^2+1)A^{20} + ([4]^2+[2]^2)A^{18} +
\Big(\frac{[5][4][3]}{[2]}+1\Big)A^{16} + ([6]^2+[2]^2)A^{14}
+ \right.\nn \\ \left.
+ \frac{[4]}{[2]}([11]+[5]+2[3])A^{12} +([6]^2+[2]^2)A^{10} + \Big(\frac{[5][4][3]}{[2]}+1\Big)A^{8}
+ \underbrace{([4]^2+[2]^2)A^{6}+ ([3]^2+1)A^4 + [2]^2A^2+1}_{
\frac{1}{(1-q^2A^2)(1-A^2)^2(1-q^{-2}A^2)}
}\right) \nn \\
\ldots \nn \\
\nn
\ee

From this we deduce the
evolution formulas for twist knots:
\be
F_1^{(m)} = A\cdot\left( -\frac{A^{2m}}{\{A\}} +\frac{1}{\{A\}}\right) \nn \\
F_2^{(m)} = qA^2\cdot\left(\frac{q^{4m}A^{4m}}{\{Aq^2\}\{Aq\}}
- \frac{[2]A^{2m}}{\{Aq^2\}\{A\}} + \frac{1}{\{Aq\}\{A\}}\right) \nn \\
{\cal F}_3^{(m)} = A^3\cdot \left( -\frac{A^{6m}}{\{Aq^2\}\{A\}\{A/q^2\}} + \frac{[3]A^{4m}}{[2]\{Aq^2\}\{A/q^2\}}
\Big(\frac{q^{4m}}{\{Aq\}} + \frac{q^{-4m}}{\{A/q\}}\Big) - \frac{[3]A^{2m}}{\{Aq^2\}\{A\}\{A/q^2\}}
+ \frac{1}{\{Aq\}\{A\}\{A/q\}}\right)\nn \\
\!\!\!\!\!\!\!\!\!\!\!\!\!\!\!\!\!\!\!\!
{\cal F}_4^{(m)} = A^4\cdot\left( \frac{A^{8m}}{\{Aq\}\{A\}^2\{A/q\}}
-\frac{[2]^2 A^{6m}}{\{Aq^2\}\{A\}^2\{A/q^2\}}
+ \frac{[3]A^{4m}(q^{4m} +q^{-4m})}{\{Aq^2\}\{Aq\}\{A/q\}\{A/q^2\}}
 - \frac{[2]^2 A^{2m}}{\{Aq^2\}\{A\}^2\{A/q^2\}}
+ \frac{1}{\{Aq\}\{A\}^2\{A/q\}}\right)
\nn
\ee
while for symmetric representations we have \cite{MMMevo}
(note that eigenvalues are now different, $(-)^iA^iq^{i(i-1)}$):
\be
F_3^{(m)} = q^3A^3\cdot \left( -\frac{q^{12m}A^{6m}}{\{Aq^4\}\{Aq^3\}\{Aq^2\}}
+ \frac{[3]q^{4m}A^{4m}}{\{Aq^4\}\{Aq^2\}\{Aq\}}
  - \frac{[3]A^{2m}}{\{Aq^3\}\{Aq^2\}\{A\}}
+ \frac{1}{\{Aq^2\}\{Aq\}\{A\}}\right)\nn \\
%\!\!\!\!\!\!\!\!
F_4^{(m)} = q^6A^4\cdot \left( \frac{q^{24m} A^{8m}}{\{Aq^6\}\{Aq^5\}\{Aq^4\}\{Aq^3\}}
-\frac{[4]q^{12m} A^{6m}}{\{Aq^6\}\{Aq^4\}\{Aq^3\} \{Aq^2\}}
+ \frac{\frac{[4][3]}{[2]}q^{4m}A^{4m} }{ \{Aq^5\}\{Aq^4\}\{Aq^2\}\{Aq\} }
- \right. \nn \\ \left.
- \frac{[4] A^{2m}}{\{Aq^4\}\{Aq^3\}\{Aq^2\}\{A\}}
+ \frac{1}{\{Aq^3\}\{Aq^2\}\{Aq\}\{A\}}\right)
\ee
Of course, these formulas can be checked for other low-intersection twist knots
$6_1,8_1,10_1$, respectively for $m =-2,-3,-4$.

\section{Factorization of differential expansion for double-braids
\label{factdb22}}

Much more interesting are the non-twist knots from the double-braid family.
Already in the first two examples we obtain:
\be
H_{[22]}^{8_3}=1- [2]^2A^{-2}Z_{[22]}^{(0)}F_1^{(2)}(A)F_1^{(-2)}(A)
+ [3]Z_{[22]}^{(0)}\Big(G_{[2]}^{8_3}(A,q)Z_{[22]}^{(1)}+ G_{[2]}^{8_3}(A,q^{-1})Z_{[22]}^{(-1)}\Big) - \nn\\
- A^{-6}[2]^2Z_{[22]}^{(1)}Z_{[22]}^{(0)}Z_{[22]}^{(-1)} {\cal F}_3^{(2)}{\cal F}_3^{(-2)}
+ A^{-8}Z_{[22]}^{(1)}\Big(Z_{[22]}^{(0)}\Big)^2 Z_{[22]}^{(-1)}{\cal F}_4^{(2)}{\cal F}_4^{(-2)}
\nn \\ \nn \\
H_{[22]}^{10_3}=1- [2]^2A^{-2}Z_{[22]}^{(0)}F_1^{(2)}(A) F_1^{(-3)}(A)
+ [3]Z_{[22]}^{(0)}\Big(G_{[2]}^{10_3}(A,q)Z_{[22]}^{(1)}+ G_{[2]}^{10_3}(A,q^{-1})Z_{[22]}^{(-1)}\Big) - \nn\\
- A^{-6}[2]^2Z_{[22]}^{(1)}Z_{[22]}^{(0)}Z_{[22]}^{(-1)} {\cal F}_3^{(2)}{\cal F}_3^{(-3)}
+ A^{-8}Z_{[22]}^{(1)}\Big(Z_{[22]}^{(0)}\Big)^2 Z_{[22]}^{(-1)}{\cal F}_4^{(2)}{\cal F}_4^{(-3)}
\ee
with $G_{[2]}^{8_3} = q^{-2}A^{-4} F_2^{(2)}F_2^{(-2)}$ and
$G_{[2]}^{10_3} = q^{-2}A^{-4} F_2^{(2)}F_2^{(-3)}$.

\bigskip

Clearly, the coefficients ${\cal F}$ factorize into products of the twist-family ${\cal F}$-factors!

Assuming that this amusing factorization is always true for {\it all} double braids, we conjecture:
\be
H_{[22]}^{(m,n)} \ \stackrel{}{=} \ 1- [2]^2A^{-2}Z_{[22]}^{(0)}\cdot F_1^{(m)}(A)F_1^{(n)}(A)
+ \nn \\
+  [3]A^{-4} Z_{[22]}^{(0)}\Big(q^{-2} Z_{[22]}^{(1)} \cdot F_2^{(m)}(A,q)F_2^{(n)}(A,q)
\ +\  q^{2} Z_{[22]}^{(-1)}\cdot F_2^{(m)}(A,q^{-1})F_2^{(n)}(A,q^{-1})\Big)
- \nn \\
- A^{-6}[2]^2Z_{[22]}^{(1)}Z_{[22]}^{(0)}Z_{[22]}^{(-1)}\cdot {\cal F}_3^{(m)}{\cal F}_3^{(n)}
\ +\  A^{-8}Z_{[22]}^{(1)}\Big(Z_{[22]}^{(0)}\Big)^2 Z_{[22]}^{(-1)}\cdot {\cal F}_4^{(m)}{\cal F}_4^{(n)}
\ee
what can be tested in the other two available examples in table (\ref{dbraidtable}) -- $7_4$ and $9_5$.
This conjecture provides a double-evolution matrix, which through (\ref{debarS}) gives the entries of
$\sqrt{d_{\bar X}d_{\bar Y}}   \bar S_{\bar X\bar Y}$.
This actually {\it proves} the conjecture, because
the result reproduces the matrix $\bar S$, {\it derived} from a very different double evolution
in \cite{mmms3}:

{\footnotesize
\be
\bar S^{[2,2]} = \frac{1}{d_{[2,2]}}\left(\begin{array}{cccccc}
\sqrt{\bar d_1} & \sqrt{\bar d_2} & \sqrt{\bar d_3} & \sqrt{\bar d_4} & \sqrt{\bar d_5} & \sqrt{\bar d_6}
\\ \\
12 & \frac{D_1D_{-1}}{[2]^2D_2D_{-2}}\gamma_1 & -\frac{\sqrt{D_{3}D_{1}}D_{-1}D_0}{[2]^2D_2D_{-2}}\gamma_2
& \frac{\sqrt{D_{-3}D_{-1}}D_0D_{1}}{[2]^2D_2D_{-2}}\gamma_3
& -\frac{\sqrt{D_{3}D_{1}D_{-1}D_{-3}}D_1D_{-1}}{[2]^2[3]D_2D_{-2}}\gamma_4 & -15
\\ \\
13 & 23 & \frac{D_0^2}{[2]^2[3]D_2D_{-2}}\gamma_5
& -\frac{\sqrt{D_{3}D_{1}D_{-1}D_{-3}}D_0^2[3]}{[2]^2D_2D_{-2}} & -24 & 14
\\ \\
14 & 24 & 34 & 33 & -23 & 13
\\ \\
15 & 25 & -24 & -23 & 22 & -12
\\ \\
16 & -15 & 14 & 13 & -12 & 11
\end{array}\right)
\nn
\ee
}

\bigskip

\noindent
with
{\footnotesize$\gamma_1=[3]D_2D_{-2} - [2]^2, \ \gamma_2=D_2D_{-3} - [2], \ \gamma_3=D_3D_{-2} - [2],
 \ \gamma_4=D_3D_{-3} - 2[3] - 1, \ \gamma_5=D_3D_2D_{-2}D_{-3} - D_{3}D_{-3} + [2]^2$}.

\bigskip

\noindent
Here  $D_i = \frac{\{Aq^i\}}{\{q\}}$,
while $\bar d_i$ are quantum dimensions of the six irreducible
representations in the product $[22]\otimes \overline{[22]}$:
{\footnotesize
\be
\!\!\!\!\!\!
\bar d_1=1, \ \ \ \ \ \ \
\bar d_2=D_1D_{-1}, \ \ \ \ \ \ \
\bar d_3=\frac{D_{3}D_{0}^2D_{-1}}{[2]^2}, \ \ \ \ \ \ \
\bar d_4=\frac{D_{1}D_{0}^2D_{-3}}{[2]^2}, \ \ \ \ \ \ \
\bar d_5=\frac{D_{3}D_{1}^2D_{-1}^2D_{-3}}{[3]^2}, \ \ \ \ \ \ \
\bar d_6=\frac{D_3D_{2}^2D_1D_{-1}D_{-2}^2D_{-3}}{[3]^2[2]^4 }
\nn
\ee}
The matrix has two symmetries: $\bar S_{i,j} = \bar S_{j,i}$ and $\bar S_{i,j} = \pm\bar S_{7-j,7-i}$,
the signs are shown explicitly in above table,
where "12" at the place of $\bar S_{21}$ means that this element is equal to $\bar S_{12}$ etc.
Presented is the matrix, obtained by our new procedure: it differs a little (in signs and order of lines
-- as allowed by conjugation freedom)
from the one in \cite{mmms3}.

\bigskip

As already mentioned, another inclusive Racah matrix $S$ is obtained from (\ref{SvsbS}),
by diagonalization of $\bar T\bar S\bar T$ with
\be
\bar T = {\rm diag} \Big( 1,   -A,     q^2 A^2, q^{-2}A^2,   -A^3,   A^4\Big)
\ee
The eigenvalues are equal to
\be
T^{-1} =  {A^4}\cdot  {\rm diag} \Big(  q^8,   -q^4,   q^2,   1/q^2,   -1/q^4,   1/q^8  \Big)
\ee
and the entries of $S_{XY}$ are then given by Cramer rule as minors of the matrix
$\bar T\bar S\bar T - T$, like in a similar procedure, described in \cite{mmms3}.

\bigskip

Factorization property holds for symmetric and antisymmetric representations as well:
for typical examples
{\footnotesize
\be
\!\!\!\!\!\!\!
H_{[3]}^{(m,n)} \ \stackrel{}{=} \ 1- [3]A^{-2}Z_{[4]}^{(0)}\cdot
 F_1^{(m)}(A)F_1^{(n)}(A)
+ \nn \\
+  [3]q^{-2}A^{-4} Z_{[3]}^{(1)} Z_{[4]}^{(0)}  \cdot   F_2^{(m)}(A,q)F_2^{(n)}(A,q)
- q^{-6}A^{-6} Z_{[4]}^{(2)}Z_{[4]}^{(1)}Z_{[4]}^{(0)}\cdot   {F}_3^{(m)}(A,q){F}_3^{(n)}(A,q)
\nn\ee
}
and
{\footnotesize
\be
H_{[1111]}^{(m,n)} \ \stackrel{}{=} \ 1- [4]A^{-2} Z_{[1111]}^{(0)}\cdot
  F_1^{(m)}(A)F_1^{(n)}(A)
%+ \nn \\
+  \frac{[4][3]}{[2]} q^2A^{-4} Z_{[1111]}^{(0)} Z_{[1111]}^{(-1)}
\cdot   F_2^{(m)}(A,q^{-1})F_2^{(n)}(A,q^{-1})
- \nn \\
-[4] q^6A^{-6} Z_{[1111]}^{(0)}Z_{[1111]}^{(-1)}Z_{[1111]}^{(-2)}\cdot
{F}_3^{(m)}(A,q^{-1}){F}_3^{(n)}(A,q^{-1})
%+ \nn \\
+ q^{12}A^{-8} Z_{[1111]}^{(0)}Z_{[1111]}^{(-1)}Z_{[1111]}^{(-2)}Z_{[1111]}^{(-3)}\cdot
 {F}_4^{(m)}(A,q^{-1}){F}_4^{(n)}(A,q^{-1})
\nn\ee
}

However, factorization is {\it not} extended to polynomials of the form
\be
H_R^{(m_1,\ldots,m_k)} =
d_R\cdot (\bar S \bar T^{2m_1}\bar S \bar T^{2m_2} \bar S \ldots \bar S\bar T^{2m_k}\bar S)_{\emptyset\emptyset}
\ee
with $k\neq 2$ -- it fails already for the fundamental representations of the simplest knots
of thus type, like $5_1=(1,1,1,1)$ and $7_3=(1,1,1,2)$.

\bigskip

Instead we claim that it {\it is} extended to other rectangular representations --
but only for {\it double} antiparallel braids.
%\be
%{\cal F}^{(m,n)}_{\A,\B} = A^{-2|\A,\B |}\cdot {\cal F}_{\A,\B}^{(m)}\cdot {\cal F}_{\A,\B}^{(n)}
%\ \ \ \ \  \  \forall\  (\A,\B), \ \ {\rm see \ notation\ below}
%\label{factprop}
%\ee
This is, however, exactly what we need to get the matrices $\bar S$.

\section{Other rectangular representations $R=[r^s]$
\label{plan}}

Now {\bf the plan} is clear.

$\bullet$
We take general differential expansion of rectangular HOMFLY for defect-zero knots
from \cite{rect41}.
It has the form
\be
H_{[r^s]}^{{\cal K}^{(0)}} = \sum_{\A,\B } C^{\A,\B }_{[r^s]}\cdot Z_{\A,\B }\cdot {\cal F}_{\A,\B}^{{\cal K}^{(0)}}
\ee
where sum goes over peculiar multi-floor pyramids, labeled by  pairs of the Young-like diagrams $\A,\B$,
All the coefficients ${\cal F}_{\A,\B}^{4_1}=1$ for the figure-eight knot
and ${\cal F}_{\A,\B}^{3_1}$ are just monomials (powers of $q$ and $A$) for the trefoil.

$\bullet$
Guess a general formula for the coefficients ${\cal F}_{\A,\B}^{(m)}$ in the case of twist knots.

$\bullet$
Assume that antiparallel double braids satisfy factorization
\be
\boxed{
{\cal F}_{\A,\B}^{(m,n)} =
%???
\frac{{\cal F}_{\A,\B}^{(m)}{\cal F}_{\A,\B}^{(n)}}
{{\cal F}_{\A,\B}^{(1)}{\cal F}_{\A,\B}^{(-1)}}
}
\label{factprop}
\ee
where ${\cal F}_{\A,\B}^{(1)} = \Big(-q^{|\A|-|\B|}A^{2}\Big)^{|\A,\B|}$ and ${\cal F}_{\A,\B}^{(-1)} =1$.

$\bullet$ Representing double-braid case ($k=2$) in the form of a double evolution in $m_1$ and $m_2$,
use (\ref{debarS}) to extract the Racah matrix $\bar S_{[r^s]}$.

$\bullet$ Diagonalizing $\bar T\bar S\bar T$ with the help of known eigenvalues and Cramer rule,
as explained in \cite{mmms3}, obtain another Racah matrix $S_{[r^s]}$.

$\bullet$ Calculate $H_{[r^s]}$ for any desired arborescent knot by the technique of \cite{mmmrv,MMfam}
(some additional care/checks can be needed with effective vertices in the multi-finger case)
and make possible checks.

The most artful part for today is the second step: revealing the structure of ${\cal F}$ for twist knots
from very special examples -- symmetric and antisymmetric representations $R=[r]$ and $R=[1^r]$
and the only double-floor example, provided by $R=[22]$.

\bigskip

More formally, complementing the conjecture of
\cite{rect41} by above suggestions, we look for the $[r^s]$-colored
HOMFLY in the following form:
\be
\!\!\!\!\!\!
\boxed{
H^{(m,n)}_{[r^s]} \stackrel{?}{=} \sum_{F=0}^{{\rm min}(r,s)}
\sum_{\{\A,\B\}} W_{\A,\B}\cdot
\boxed{
\frac{{\cal F}_{\A,\B}^{(m)}{\cal F}_{\A,\B}^{(n)}}
{{\cal F}_{\A,\B}^{(1)}{\cal F}_{\A,\B}^{(-1)}}
} \cdot
\left(\prod_{f=1}^{F}
%\boxed{(-A^2)^{p_f} q^{p_f(a_f- b_f)}}\cdot
\Big(C_{a_f+b_f}^{b_f}\Big)^2 C_{r+b_f}^{p_f} C_{s+a_f}^{p_f}
\right)\cdot
{\footnotesize
{{{{
%{\ldots\phantom{\oint^{\oint^5_5}_{\oint^5_5}}
%\over
\boxed{a_F\ldots 0 \ldots -b_F}
%}
\over  \ldots\phantom{\oint^{\oint^5_5}_{\oint^5_5}}  }
\over\boxed{a_3\ \ \ \ \ldots \ \ \ \ 1\, 0\,{-1}\ \ \ldots \ \ \ \ -b_3}  }
\over\boxed{a_2\ \ \ \ \ \ \ldots  \ \ \ \ 1\, 0\,{-1}\ \  \ \ \ldots\ \ \  \ \ -b_2}  }
\over\boxed{a_1\ \ \ \ \ \ \   \ \ldots \ \ \ \ \ \ \ 1\, 0\,{-1}\ \ \ \ \ \ \ \ldots\ \ \ \ \ \ \ \ -b_1}}
}
=}
%\label{H41rspict}
\nn
\ee
\be
=1+ \sum_{F=1}^{{\rm min}(r,s)} \sum_{
\stackrel{0\leq a_F <\ldots < a_3<a_2<a_1<r}{0\leq b_F < \ldots < b_3< b_2<b_1<s}}
\ \ \prod_{f'<f''}^F
%W\{a_{f'},b_{f'}|a_{f''},b_{f''}\}
\left(\frac{[a_{f'}-a_{f''}][b_{f'}-b_{f''}]}{[a_{f'}+b_{f''}+1][a_{f''}+b_{f'}+1]}\right)^2
\cdot
\boxed{
\frac{{\cal F}_{\A,\B}^{(m)}{\cal F}_{\A,\B}^{(n)}}
{{\cal F}_{\A,\B}^{(1)}{\cal F}_{\A,\B}^{(-1)}}
}
\cdot
\label{41rs}
\ee
\vspace{-0.2cm}
\be
\cdot \prod_{f=1}^F \left(
\left(\frac{[a_f+b_f]!}{([a_f]![b_f]! }\right)^2
\frac{[r+b_f]![s+a_f]!}{[r-1-a_f]![s-1-b_f]!\big([a_f+b_f+1]!\big)^2}
 \prod_{i_f=-b_f}^{a_f} \{Aq^{i_f+r}\}\{Aq^{i_f-s}\}\right)
 \nn
\ee
and our {\it new goal} is to associate the evolution functions
${\cal F}_{\A,\B}^{(m)}$ with every non-empty pyramid
\be
\{\A,\B\}
= \{r>a_1>a_2\ldots>a_f\geq 0\}\cup \{s>b_1>b_2>\ldots>b_f\geq 0\}
\ee
These functions depend on the eigenvalues $\lambda_c$,
which are powers of $A$ and $q$.
%and their full quantity for the case of $R=[r^s]$ is the maximal value
%of the sum over floors $f$ of the pyramid
%\be
%???
%{\rm maximum}_{\{\A,\B\}\in R}\Big(\sum_{f=1}^F (a_f+b_f+2)\Big) =
%{\rm maximum}_{\{\A,\B\}\in R}\Big(F + \sum_{f=1}^F (a_f+b_f+1)\Big) = {\rm min}(r,s) + r^s
%\ee
However, for concrete pyramid $\{\A,\B\}$ the function ${\cal F}_{\A,\B}^{(m)}$
depends on only {\it some} of these eigenvalues -- and this subset depends on the pyramid and
not on $r$ and $s$ (these parameters, however, select particular pyramides that contribute
to $H_{[r^s]}$).
Schematically,
\be
{\cal F}_{\A,\B}^{(m)} = \sum_{c\in I_{\A,\B}} \lambda_c^{2m} \cdot \frac{\xi_c}{\prod_{i\in J_{\A,\B}}
\{Aq^i\}}
\ee
The first two tasks are to describe the sets $I_{\A,\B}$ and $J_{\A,\B}$, and the third is
the finding of the combinatorial coefficients $\xi_c$.
In this paper they are fulfilled only partly -- for the single-floor diagrams $\{\A,\B\}$.

\section{On the structure of ${\cal F}_{\A,\B}^{(m)}$
\label{Ftwist}}

\subsection{Already known cases}

To begin with let us remind this structure in the case of {\bf symmetric representations} $[r]$,
where the relevant diagrams are just single-floor boxes $\boxed{a\ \ldots\ 0}$ with $b=0$.
Then, from \cite{MMMevo} we know the answer:
\be
{\cal F}_{_{\tiny{\boxed{a\ \ldots\ 0}}}}^{(m)} =q^{\frac{a(a+1)}{2}}A^{a+1}\sum_{c=0}^{a+1}
\frac{(-)^c[a+1]!}{[c]![a+1-c]!}
  \cdot\lambda_c^{2m} \cdot \frac{\{Aq^{2c-1}\}}
{\prod_{i=c-1}^{c+a} \{Aq^i\}}
\label{Fsym}
\ee
It depends on $a+2$ eigenvalues (of the quantum ${\cal R}$-matrix)  $\lambda_c = (-)^c q^{c(c-1)}A^c$.
Likewise, for the {\bf antisymmetric representations} $[1^s]$
contributing are only the boxes $\boxed{0\ \ldots\ -b}$ with $a=0$, and
\be
{\cal F}_{_{\tiny{\boxed{0\ \ldots\ -b}}}}^{(m)} =
q^{-\frac{b(b+1)}{2}}A^{b+1} \sum_{c=0}^{b+1}
  \frac{(-)^c[b+1]!}{[c]![b+1-c]!} \cdot \lambda_{-c}^{2m} \cdot \frac{\{Aq^{1-2c}\}}
{\prod_{i=c-1}^{c+b} \{Aq^{-i}\}}
\label{Fasym}
\ee
with $b+2$ eigenvalues $\lambda_{-c} = (-)^c q^{-c(c-1)} A^c$.
In fact, it is already a challenge to unify
%these two formulas
(\ref{Fsym}) and (\ref{Fasym}) into a single formula.

\bigskip

The knowledge of {\bf the $R=[22]$ case} adds to
\be
{\cal F}_{_{\tiny \boxed{0} }}^{(m)} = A\cdot \left(\lambda_0^{2m}\, \frac{\{A/q\}}{\{A/q\}\{A\}}
- \lambda_1^{2m}\, \frac{\{Aq\}}{\{A\}\{Aq\}}\right) = A\cdot \frac{\lambda_0^{2m}-\lambda_1^{2m}}{\{A\}}
= \frac{1-A^{2m}}{1-A^{-2}} \nn \\ \nn \\
{\cal F}_{_{\tiny\boxed{1 \ 0 }}}^{(m)} =
qA^2 \cdot \left(\lambda_0^{2m}\, \frac{\{A/q\}}{\{A/q\}\{A\}\{Aq\}}
- [2]\,\lambda_1^{2m}\, \frac{\{Aq\}}{\{A\}\{Aq\}\{Aq^2\}}
+ \lambda_2^{2m}\,\frac{\{Aq^3\}}{\{Aq\}\{Aq^2\}\{Aq^3\}} \right) \nn \\ \nn \\
{\cal F}_{_{\tiny\boxed{0\  -1}}}^{(m)} =
q^{-1}A^2 \cdot \left(\lambda_0^{2m}\, \frac{\{Aq\}}{\{Aq\}\{A\}\{A/q\}}
- [2]\,\lambda_1^{2m}\, \frac{\{A/q\}}{\{A\}\{A/q\}\{A/q^2\}}
+ \lambda_{-2}^{2m}\,\frac{\{A/q^3\}}{\{A/q\}\{A/q^2\}\{A/q^3\}} \right)
\label{F22a}
\ee
two more pyramids:
\be
\!\!\!\!\!\!
{\cal F}_{_{\tiny\boxed{1\ 0\  -1}}}^{\,(m)} =
A^3\cdot \left(
\frac{\lambda_0^{2m}}{\{Aq\}\{A\}\{A/q\}}
- \frac{[3]\lambda_1^{2m}}{\{Aq^2\}\{A\}\{A/q^2\}}
+ \frac{[3]\lambda_2^{2m}}{[2]\{Aq^2\}\{Aq\}\{A/q^2\}}
+ \frac{[3]\lambda_{-2}^{2m}}{[2]\{Aq^2\}\{A/q\}\{A/q^2\}}
- \right.\nn \\ \nn \\ \left.
-\frac{\lambda_{03}^{2m}}{\{Aq^2\}\{A\}\{A/q^2\}}
\right) \ \ \ \ \ \ \ \ \ \ \ \ \ \ \ \ \ \ \ \ \ \ \ \ \ \ \ \ \ \ \ \ \ \ \ \
\ \ \ \ \ \ \ \ \ \ \ \ \ \ \ \ \ \ \nn \\ \nn \\ \nn \\
\!\!\!\!\!\!
{\cal F}_{_{\tiny\!\!\!\!\!\!{{\boxed{0}\ \ \ \ }\over{\boxed{1\ 0\ -1}}}}}^{\,(m)} =
 A^4\cdot\left( \frac{\lambda_0^{2m}}{\{Aq\}\{A\}^2\{A/q\}}
- \frac{[2]^2\, \lambda_1^{2m}}{\{Aq^2\}\{A\}^2\{A/q^2\}}
+ \frac{[3]\,\lambda_{2}^{2m} }{\{Aq^2\}\{Aq\}\{A/q\}\{A/q^2\}}
+ \frac{[3]\,\lambda_{-2}^{2m} }{\{Aq^2\}\{Aq\}\{A/q\}\{A/q^2\}}
-\right.\nn
\ee
\vspace{-0.5cm}
\be \left. \ \ \ \ \ \ \ \ \ \ \ \ \ \ \ \ \ \ \ \ \ \ \ \ \ \ \ \ \ \ \ \ \ \ \ \
-\frac{[2]^2\, \lambda_{03}^{2m}}{\{Aq^2\}\{A\}^2\{A/q^2\}}
 +\frac{\lambda_{04}^{2m}}{\{Aq\}\{A\}^2\{A/q\}}
\right)
\label{F22b}
\ee
with five and six eigenvalues, associated with the six irreducible representations in the
product $[22]\otimes\overline{[22]}$:
\be
\lambda_0 = 1, \ \ \lambda_1 = -A =\lambda_{-1}, \ \ \lambda_{\pm 2} = q^{\pm 2}A^2, \ \
\lambda_{03} = -A^3, \ \ \lambda_{04}=A^4
\label{ev22}
\ee

\subsection{The first unknown case:  $R=[33]$}

Now there are ten irreps in $[33]\otimes\overline{[33]}$ and ten eigenvalues:
six "old" ones, the same as (\ref{ev22}),
and four "new":
\be
\lambda_{+3} = -q^6A^3,   \ \ \lambda_{+_04}  =q^4A^4, \ \ \lambda_{+_05} = -q^4A^5,
\ \ \lambda_{+_06} = q^6A^6
\label{ev33}
\ee
(somewhat strange labels in $\lambda_{+_0c}$ emphasize that the powers of $q$ are smaller
than $c(c-1)$ in $\lambda_{+c}$, familiar from symmetric representation case (\ref{Fsym}),
see s.\ref{eigenv} below for a more systematic description).
The "old" pyramids are described by the same formulas (\ref{F22a}) and (\ref{F22b}),
involving only "old" eigenvalues (\ref{ev22}).
Of the four "new" pyramids one is still described by (\ref{Fsym}):
\be
{\cal F}_{\tiny\boxed{2\ 1 \ 0 }}^{(m)} =
q^3A^3\cdot \left(\lambda_0^{2m}\, \frac{\{A/q\}}{\{A/q\}\{A\}\{Aq\}\{Aq^2\}}
- [3]\,\lambda_1^{2m}\, \frac{\{Aq\}}{\{A\}\{Aq\}\{Aq^2\}\{Aq^3\}}
+ \right. \nn \\ \left.
+ [3]\,\lambda_2^{2m}\, \frac{\{Aq^3\}}{\{Aq\}\{Aq^2\}\{Aq^3\}\{Aq^4\}}
- \lambda_3^{2m}\, \frac{\{Aq^5\}}{\{Aq^2\}\{Aq^3\}\{Aq^4\}\{Aq^5\}}
\right)
\ee
while three more we should guess:
\be
{\cal F}_{\tiny\boxed{2\ 1 \ 0 \ -1}}^{(m)} \ \stackrel{?}{=} \
q^2A^4\cdot\left(
\frac{\lambda_0^{2m}}{\{Aq^2\}\{Aq\}\{A\}\{A/q\}}
- \frac{[4]\,\lambda_1^{2m}}{\{Aq^3\}\{Aq^2\}\{A\}\{A/q^2\}}
+ \right.\nn \\ \nn \\ \left.
+ \frac{[4] \lambda_2^{2m}  }{\{Aq^4\}\{Aq^2\}\{Aq\}\{A/q^2\}}
+ \frac{\frac{[4]}{[2]} \lambda_{-2}^{2m} }{\{Aq^3\}\{Aq^2\}\{A/q\}\{A/q^2\}}
\underline{
- \frac{\frac{[4]}{[3]} \lambda_{+3}^{2m}}{\{Aq^4\}\{Aq^3\}\{Aq^2\}\{A/q^2\}}
-\frac{\frac{[4][2]}{[3]} \lambda_{03}^{2m}}{\{Aq^4\}\{Aq^2\}\{A\}\{A/q^2\}}
\ }
+ \right.
\nn
\ee
\vspace{-0.2cm}
\be
 \left.
+ \frac{\lambda_{+_04}^{2m}}{\{Aq^4\}\{Aq^3\}\{Aq\}\{A/q^2\}}
\right)
\label{F2101}
\ee

\be
{\cal F}_{\tiny\!\!\!\!\!\!\!\!{{\boxed{0}\ \ }\over{\boxed{2\ 1\ 0\ -1}}}}^{(m)}  \ \stackrel{?}{=} \
q^2A^5\cdot\left(
\frac{\lambda_0^{2m}}{\{Aq^2\}\{Aq\}\{A\}^2\{A/q\}}
- \frac{[4]\,\lambda_1^{2m}}{\{Aq^3\}\{Aq^2\}\{A\}^2\{A/q^2\}}
- \frac{ \lambda_1^{2m}}{\{Aq^3\}\{Aq\}\{A\}^2\{A/q^2\}}
+ \right.\nn \\ \left.
\!\!\!\!\!\!\!\!\!\!\!\!\!\!\!\!\!\!\!\!\!\!\!\!\!\!\!\!\!\!\!\!\!\!\!\!
+ \frac{\frac{[4][3]}{[2]} \lambda_2^{2m}  }{\{Aq^4\}\{Aq^2\}\{Aq\}\{A/q\}\{A/q^2\}}
+ \frac{[4] \lambda_{-2}^{2m}  }{\{Aq^3\}\{Aq^2\}\{Aq\}\{A/q\}\{A/q^2\}}
- \right.\nn \\ \left.
- \frac{\frac{[4]}{[2]} \lambda_{+3}^{2m}}{\{Aq^4\}\{Aq^3\}\{Aq^2\}\{A/q\}\{A/q^2\}}
-\frac{[4][2] \lambda_{03}^{2m}}{\{Aq^4\}\{Aq^2\}\{A\}^2\{A/q^2\}}
+ \right.\nn \\ \left.
\!\!\!\!\!\!\!\!\!\!\!\!\!\!\!\!\!\!\!\!\!\!\!\!
+ \frac{[3] \lambda_{+_04}^{2m}}{\{Aq^4\}\{Aq^3\}\{Aq\}\{A\}\{A/q^2\}}
+ \frac{\frac{[4]}{[2]}\lambda_{04}^{2m}}{\{Aq^4\}\{Aq\} \{A\}^2\{A/q\}}
- \frac{\lambda_{+_05}^{2m}}{\{Aq^4\}\{Aq^2\}\{Aq\}\{A\}\{A/q\}}
\right)
\label{21010}
\ee

\bigskip

\be
{\cal F}_{\tiny\!\!\!\!{{\boxed{1\ 0}\ \ \ \ }\over{\boxed{2\ 1\ 0\ -1}}}}^{ (m)}
 \ \stackrel{?}{=} \
q^3A^6\cdot\left(
\frac{\lambda_0^{2m}}{\{Aq^2\}\{Aq\}^2\{A\}^2\{A/q\}}
- \frac{[3][2]\,\lambda_1^{2m}}{\{Aq^3\}\{Aq^2\}\{Aq\}\{A\}^2\{A/q^2\}}
+ \right.\nn \\ \left.
\!\!\!\!\!\!\!\!\!\!\!\!\!\!\!\!\!\!\!\!\!\!\!\!\!\!\!\!\!\!\!\!\!\!\!\!
+ \frac{[3]^2 \lambda_2^{2m}  }{\{Aq^4\}\{Aq^2\}\{Aq\}^2\{A/q\}\{A/q^2\}}
+ \frac{\frac{[4][3]}{[2]} \lambda_{-2}^{2m}  }{\{Aq^3\}\{Aq^2\}^2\{Aq\}\{A/q\}\{A/q^2\}}
-\right.\nn\\ \left.
- \frac{[4] \lambda_{+3}^{2m}}{\{Aq^4\}\{Aq^3\}\{Aq^2\}\{A\}\{A/q\}\{A/q^2\}}
-\frac{[4][2]^2 \lambda_{03}^{2m}}{\{Aq^4\}\{Aq^2\}^2 \{A\}^2\{A/q^2\}}
+ \right.\nn \\ \left.
\!\!\!\!\!\!\!\!\!\!\!\!\!\!\!\!\!\!\!\!\!\!\!\!\!\!\!\!\!\!\!\!\!\!\!\!
+ \frac{[3]^2 \lambda_{+_04}^{2m}}{\{Aq^4\}\{Aq^3\}\{Aq\}^2\{A\}\{A/q^2\}}
+ \frac{\frac{[4][3]}{[2]}\lambda_{04}^{2m}}{\{Aq^4\}\{Aq^3\}\{Aq\}\{A\}^2\{A/q\}}
%\cdot\frac{\{Aq^2\}\{A/q^2\}}{\{Aq\}\{A/q\}}
- \right.\nn \\ \left.
-\frac{[3][2]\lambda_{+_05}^{2m}}{\{Aq^4\} \{Aq^2\}^2\{Aq\}\{A\}\{A/q\}}
%\cdot\frac{\{Aq^3\}\{A/q^2\}}{\{Aq^2\}\{A/q\}}
+ \frac{\lambda_{+_06}^{2m}}{\{Aq^3\}\{Aq^2\}^2\{Aq\}^2\{A\} }
%\cdot\frac{\{Aq^4\}\{A/q^2\}}{\{Aq^2\}\{A/q\}}
\right)
\label{210110}
\ee

\bigskip

\noindent
Such guesses are motivated by a number of  requirements:

\bigskip

$\bullet$ similarity to (\ref{Fsym}), (\ref{Fasym}) and (\ref{F22b}),
implying that ${\cal F}_{\A,\B}$ is a sum of powers $\lambda_c^{2m}$ with nicely factorized coefficients,

$\bullet$
observation in these examples of certain regularity in the positions of poles,
coming from $\{Aq^i\}$ with $-2b<i<2a$ in denominators, for terms with different $\lambda_c^{2m}$,

$\bullet$ requirement that each term in the sum over $\lambda_c$ depends on {\it even} powers $q$,

$\bullet$ requirement, that entire ${\cal F}_{\A,\B}^{(m)}$ is a polynomial for any $m$,
i.e. all poles cancel after summation over $c$,

$\bullet$ vanishing ${\cal F}_{\A,\B}^{(0)}=0$ at $m=0$, i.e. for the unknot,

$\bullet$ ${\cal F}_{\A,\B}^{(-1)}=1$ for $m=-1$, i.e. for the figure eight knot, see \cite{rect41},

$\bullet$ ${\cal F}_{\A,\B}^{(1)} = \prod_f (-q^{a_f-b_f}A^2)^{a_f+b_f+1}$ for $m=1$,
i.e. for the trefoil, see \cite{rect41}.

\bigskip

\noindent
For the particular case of $R=[33]$ these are rather restrictive requirements
and the above guess is actually less ambiguous, than it can seem.
In any case, the real confirmation   comes  {\it a posteriori} -- from the final expressions for
Racah matrices and reasonable answers they provide for the $[33]$-colored knot polynomials.

\bigskip

With these guesses we can calculate
\be
H_{[33]}^{(m,n)}\ \stackrel{?}{=}\ 1 - [3][2]\,    Z_{[33]}^{(0)}
\cdot A^{-2}{\cal F}_{_{\tiny \boxed{0} }}^{(m)} {\cal F}_{_{\tiny \boxed{0} }}^{(n)}
+\nn \\
+  \Big([3]^2   Z_{[33]}^{(+1)}
\cdot q^{-2}A^{-4}{\cal F}_{_{\tiny\boxed{1 \ 0 }}}^{(m)}{\cal F}_{_{\tiny\boxed{1 \ 0 }}}^{(n)}
+ \frac{[3][4]}{[2]}  Z_{[33]}^{(-1)}
\cdot q^2A^{-4} {\cal F}_{_{\tiny\boxed{0\  -1}}}^{(m)}{\cal F}_{_{\tiny\boxed{0\  -1}}}^{(n)}\Big)
Z_{[33]}^{(0)}
-\nn\\
-  \Big([4]  \, Z_{[33]}^{(+2)}
\cdot q^{-6}A^{-6}{\cal F}_{\tiny\boxed{2\ 1 \ 0 }}^{(m)}\,{\cal F}_{\tiny\boxed{2\ 1 \ 0 }}^{(n)}
+ [4][2]^2 \,Z_{[33]}^{(-1)}
\cdot A^{-6}{\cal F}_{_{\tiny\boxed{1\ 0\  -1}}}^{\,(m)}\, {\cal F}_{_{\tiny\boxed{1\ 0\  -1}}}^{\,(n)}
\Big) Z_{[33]}^{(+1)}Z_{[33]}^{(0)}
+ \nn \\
+\Big([3]^2 \, Z_{[33]}^{(+2)}
\cdot q^{-4}A^{-8}{\cal F}_{\tiny\boxed{2\ 1 \ 0 \ -1}}^{(m)}\, {\cal F}_{\tiny\boxed{2\ 1 \ 0 \ -1}}^{(n)}
+ \frac{[3][4]}{[2]} \,Z_{[33]}^{(0)}
\cdot A^{-8}{\cal F}_{_{\tiny\!\!\!\!\!\!{{\boxed{0}\ \ \ \ }\over{\boxed{1\ 0\ -1}}}}}^{\,(m)}\,
{\cal F}_{_{\tiny\!\!\!\!\!\!{{\boxed{0}\ \ \ \ }\over{\boxed{1\ 0\ -1}}}}}^{\,(n)}\Big)
Z_{[33]}^{(+1)}Z_{[33]}^{(0)}Z_{[33]}^{(-1)}
-\nn \\
- [3][2]  Z_{[33]}^{(+2)} Z_{[33]}^{(+1)}\big(Z_{[33]}^{(0)}\big)^2 Z_{[33]}^{(-1)}
\cdot q^{-4}A^{-10} {\cal F}_{\tiny\!\!\!\!\!\!\!\!{{\boxed{0}\ \ }\over{\boxed{2\ 1\ 0\ -1}}}}^{(m)}\,
{\cal F}_{\tiny\!\!\!\!\!\!\!\!{{\boxed{0}\ \ }\over{\boxed{2\ 1\ 0\ -1}}}}^{(n)}
+ \nn \\
+Z_{[33]}^{(+2)} \big(Z_{[33]}^{(+1)} Z_{[33]}^{(0)}\big)^2 Z_{[33]}^{(-1)}
\cdot q^{-6} A^{-12} {\cal F}_{\tiny\!\!\!\!{{\boxed{1\ 0}\ \ \ \ }\over{\boxed{2\ 1\ 0\ -1}}}}^{ (m)}\,
{\cal F}_{\tiny\!\!\!\!{{\boxed{1\ 0}\ \ \ \ }\over{\boxed{2\ 1\ 0\ -1}}}}^{ (n)}
\ee
for double braids and extract the $10\times 10$ matrix $\bar S^{[33]}$,
which is described in the Appendix to this paper.
Using this matrix we can calculate $H_{[3,3]}$ for {\it some} arborescent knots,
which can be made without the use of the second exclusive matrix $S^{[33]}$.
In examples these polynomials are consistent with available Vassiliev invariants and pass other
checks from the list in \cite{mmms1}.
Building of $S$ from $\bar S$ with the help of (\ref{SvsbS})
and thus extension to arbitrary arborescent knots by the method of \cite{mmmrv,MMfam}
is also straightforward.
%but this exercise is a kind of orthogonal to the {\it ideas}, exposed in the present paper,
%therefore it will be described elsewhere.

\bigskip

For further generalizations to other rectangular representations
we need to look at the above requirements for ${\cal F}$ a little closer.
Since it is still a guesswork, it is not really formalized -- thus we provide just some sketchy
comments, followed by new conjectures.

\subsection{On nullification for the unknot}

Identities, necessary for nullification of ${\cal F}$ for the unknot, i.e.at $m=0$
are rather simple and already the first examples reveal their general structure:
if $D_a=\{Aq^a\}/\{q\}=[N+a]$, then
\be
\boxed{1\ 0} & D_2 - [2]\,D_1+D_0 = 0  \nn \\ \nn \\
\boxed{2\ 1\ 0} &  D_4D_3 -[3]\,D_4D_1 + [3]D_3D_0 - D_1D_0 = 0 \nn\\
%= D_4(D_3-q^2D_1)+\frac{1}{q^2}(D_3-q^2D_1)D_0 +[2]\left(qD_3D_0-\frac{1}{q}D_4D_1\right)= 0 \nn \\
\boxed{1\ 0 \ -1} & D_2D_{-2} - (\underbrace{[3]+1}_{[2]^2})\,D_1D_{-1}
+\underbrace{\frac{[3]}{[2]}D_0D_{-1}+\frac{[3]}{[2]}D_1D_0 }_{[3]\,D_0^2} = 0
\label{idsunknot}
\ee
\be
\boxed{3\ 2\ 1\ 0} & D_6D_5D_4-[4]D_6D_5D_1 + \frac{[4][3]}{[2]}D_6D_3D_0 - [4]D_5D_1D_0 + D_2D_1D_0 = 0\nn\\
\!\!\!\!\!\!\!\!\!
\boxed{2\ 1 \ 0 \ -1} & D_4D_3D_{-2} - [4] D_4D_1D_{-1}+[4]D_3D_0D_{-1}+\frac{[4]}{[2]}D_4D_1D_0
\underbrace{-\frac{[4]}{[3]}D_1D_0D_{-1}-\frac{[4][2]}{[3]}D_3D_1D_{-1}}_{-[4]D_2D_1D_{-1}}
+D_2D_0D_{-1}=0
\nn \\
\ldots \nn
\ee
They are the first of the necessary ones for polynomiality of ${\cal F}$ at all $m$ --
for that purpose above combinations should be proportional to the product of $D$-factors
in denominators of ${\cal F}$, but these products have a higher degree in $A$, thus
the proportionality coefficient is just zero.

Identities (\ref{idsunknot}) are linear combinations of
\be
D_{a+b}+D_{a-b} = \frac{[2b]}{[b]}D_a =(q^b+q^{-b})D_a \nn \\
\[b+c][b-c]\,D_aD_{-a} +[c+a][c-a]\,D_bD_{-b}+ [a+b][a-b]\,D_cD_{-c} =0  \nn \\
\ldots
\ee
which are the $q$-deformations of identities

\be
\sum_{i=1}^3 (N+a_i)\cdot (a_{i+1}-a_{i+2})=
- \sum_{i=1}^3 (N+a_i)\cdot \det\left(\begin{array}{cc} 1 & 1 \\ a_{i+1} &a_{i+2}
\end{array}\right) = 0, \nn \\
\sum_{i=1}^4 (-)^i\cdot (N+a_i)(N+b_i) \cdot \det\left(\begin{array}{ccc} 1 & 1 & 1 \\
a_{i+1}+b_{i+1}&a_{i+2}+b_{i+2} & a_{i+3}+b_{i+3} \\
a_{i+1}\cdot b_{i+1}&a_{i+2}\cdot b_{i+2} & a_{i+3}\cdot b_{i+3}
\end{array}\right) = 0
\nn \\ \nn \\
\ldots \ \ \ \ \ \ \ \ \ \ \ \ \ \ \ \ \ \ \ \nn \\ \nn \\
\sum_{i=1}^{n+2}  (-)^{(n+1)i}\underbrace{\prod_{j=1}^n (N+a_{j}^{(i)})}_{N^n+\sum_{k=1}^n N^{n-k}\mu_{k}^{(i)}}
\cdot \det_{(n+1)\times(n+1)} \left(\begin{array}{cccccc}
1&1&1&\ldots& 1 \\
\mu_{1}^{(i+1)} & \mu_1^{(i+2)} & \mu_1^{(i+3)} & \ldots &\mu_1^{(i+n+1)}\\
\mu_{2}^{(i+1)} & \mu_2^{(i+2)} & \mu_2^{(i+3)} & \ldots &\mu_2^{(i+n+1)}\\
\ldots \\
\mu_{n}^{(i+1)} & \mu_n^{(i+2)} & \mu_n^{(i+3)} & \ldots &\mu_n^{(i+n+1)}
\end{array}\right) = 0
\label{detrels}
\ee
where $\mu^{(i+n+1)}=\mu^{(i)}$.
These follow from the vanishing of $\det_{(n+2)\times(n+2)} P_i(x_{i'})$ for {\it any}
$n+2$ polynomials of degree $n$ -- because this determinant has degree $n$ in each variable $x_i$,
while Vandermonde $\prod_{i'<i''}^{n+2} (x_{i'}-x_{i''})$, to which it is obviously proportional,
has bigger degree $n+1$.
For our purposes we need peculiar collections of shifts $a_j^{(i)}$, when determinants in
(\ref{detrels}) are actually binomial coefficients.

Note, however, that nullification of ${\cal F}$ for the unknot is not by itself enough restrictive.
For example, as indicated in (\ref{idsunknot}),
it would allow to change the two underlined terms in (\ref{F2101}) for a single
$-\frac{[4]\,\lambda_{03}^{2m}}{\{Aq^4\}\{Aq^3\}\{A\}\{A/q^2\}}$ -- this, however, would give a wrong
(and in fact, non-polynomial) answer already for the trefoil $3_1$.

\subsection{Eigenvalues
\label{eigenv}}

In general the eigenvalues are parameterized as
\be
\lambda_{I,J} =  \prod_{f=1}^F (-q^{i_f-j_f}A)^{i_f+j_f+1}
\label{eigenvalues}
\ee
with $0\leq i_f\leq a_f$ and  $0\leq j_f\leq  b_f$
and an additional embedding constraint
\be
i_{f+1}<i_f\leq a_f, \ \ \ j_{f+1}<j_f\leq b_f
\label{embedev}
\ee
Also some upper floors $f$ can be left empty, i.e. allowed is also $i_f+j_f+1=0$.
In particular, $\lambda_0=1$ is associated with all empty floors,
thus it is more natural to call it $\lambda_\emptyset$.
In this sense all eigenvalues are associated with the
ordered-by-(\ref{embedev})
collection of boxes in the pyramids --
two per each  floor $f$, except for degenerate cases
when $i_f+j_f=0$ or $i_f+j_f+1=0$,
and the corresponding boxes coincide or are just absent.
This implies an obvious change of notation for the eigenvalue labeling.

Generic eigenvalue is labeled by a new pyramid:
$\lambda_{{\cal I},{\cal J}} =
\lambda_{\stackrel{}{\stackrel{i_f\ j_f}{\stackrel{\ldots}{i_1\ \ \ j_1}}}}$,
and its contribution to ${\cal F}_{\A,\B}$ contains a product of inverse differentials
$D_k^{-1}\sim \{Aq^k\}^{-1}$, which only slightly depends on $\A$ and $\B$, and a combinatorial factor,
accounting for the embedding of pyramids $\{{\cal I},{\cal J}\}\subset \{\A,\B\}$.

In these pyramid notation the eigenvalues (\ref{ev22}) and (\ref{ev33}) fit into the
big tower:
\be
\lambda_\emptyset = \lambda_0 = 1, \nn\\
\lambda_{\, 0\ 0}= \lambda_1 =\lambda_{-1} =  -A , \nn\\
\lambda_{\, 1\ 0} = \lambda_{2} =  q^{2}A^2, \ \   \lambda_{\, 0\ 1} = \lambda_{-2} =  q^{-2}A^2, \nn \\
\lambda_{\,2\ 0} = \lambda_{+3} = -q^6A^3, \ \
\lambda_{\,1\ 1} = \lambda_{03} = -A^3, \ \
\lambda_{\,0\ 2} = \lambda_{-3} = -q^{-6} A^3 \nn \\
\lambda_{\,3\ 0 } = q^{12}A^4, \ \ \lambda_{\,2\ 1}= \lambda_{+_04}  =q^4A^4, \ \
\lambda_{\stackrel{0\ 0}{1\ \ 1}}=\lambda_{04}=A^4, \ \
\lambda_{\,1\ 2}=   q^{-4}A^4, \ \  \lambda_{\,0\ 3 } = q^{-12}A^4 \nn \\
\lambda_{\,4\ 0} = -q^{20}A^5, \ \ \lambda_{\,3\ 1} = -q^{10}A^5, \ \
 \lambda_{\stackrel{0\ 0}{2\ \ 1}} = \lambda_{+_05} = -q^4A^5, \ \
 \lambda_{\,2\ 2} = -A^5, \ \
 \lambda_{\stackrel{0\ 0}{1\ \ 2}} =   -q^{-4}A^5, \ \ \ldots \nn \\
%  \lambda_{\,1\ 3} = -q^{-10}A^5, \ \ \lambda_{\,0\ 4} = -q^{-20}A^5,\nn \\
\lambda_{\,5\ 0} = q^{30}A^6, \ \ \ \ \ \ \ \   \ \ \ldots
\ \ \ \ \ \ \ \ \ \ \ \ \ \ \
\lambda_{\stackrel{1\ 0}{2\ \ 1}}= \lambda_{+_06} = q^6A^6,\ \ \ \ \ \ \ \ \ \ \ \ \ \ \ \ \ \ \ldots
\ \ \ \ \ \ \ \ \ \ \ \ \ \ \ \ \ \ \ \ \ \ \ \ \ \ \ \ \ \ \  \nn \\
\ldots
\ee

\subsection{
Single-floor pyramids
}
%Symmetric case}

Looking at a few explicit examples that we already possess,
one can {\it assume} the following structure for denominators,
associated with particular eigenvalues:
\be
\!\!\!\!\!\!\!\!\!\!\!\!\!\!
\frac{\lambda_{\emptyset}^{2m}}{\prod_{(i,j)\in (\A,\B)} \{Aq^{i-j}\}}
\ \ \Longrightarrow \ \
\frac{\lambda_{\emptyset}^{2m}}{\underd{\{Aq^a\}\ldots \{Aq\}\{A\}\{A/q\}\ldots \{A/q^b\}}}
\nn \\ \nn \\
\left\{ \begin{array}{c}
 \frac{\lambda_{0,0}^{2m}}
{\underd{\{Aq^{a+1}\} \ldots \{Aq^2\}}\  \ \{A\}\ \ \underd{\{A/q^2\}\ldots\{A/q^{b+1}\}}}
\nn \\ \nn \\
  \frac{\lambda_{1,0}^{2m}}
{\underd{\{Aq^{a+2}\} \ldots \{Aq^4\}}\  \ \underd{\{Aq^2\}\{Aq\}}\ \
\underd{\{A/q^2\}\ldots\{A/q^{b+1}\}}}
\nn \\ \nn \\
  \frac{\lambda_{2,0}^{2m}}
{\underd{\{Aq^{a+3}\} \ldots \{Aq^6\}}\ \ \underd{\{Aq^4\}\{Aq^3\}\{Aq^2\}} \  \
\underd{\{A/q^2\}\ldots\{A/q^{b+1}\}}}
\nn \\
\ldots \nn \\
\Longrightarrow \ \
\frac{\lambda_{i,0}^{2m}}
{\underd{\{Aq^{a+i+1}\}\ldots\{Aq^{2i+2}\}}\ \ \underd{\{Aq^{2i}\}\ldots\{Aq^i\}}\ \
\underd{\{A/q^2\}\ldots \{A/q^{b+1}\}}}
\end{array} \right.
  \nn
\ee
\be
  \nn \\
\left\{ \begin{array}{c}
  \frac{\lambda_{0,1}^{2m}}
{\underd{\{Aq^{a+1}\} \ldots \{Aq^2\}}\ \ \underd{\{A/q\} \{A/q^2\}}\ \
\underd{\{A/q^4\} \ldots\{A/q^{b+2}\}}}
\nn \\ \nn \\
\ldots \nn \\ \nn \\
\Longrightarrow \ \ \frac{\lambda_{0,j}^{2m}}
{\underd{\{Aq^{a+1}\}\ldots\{Aq^{2 }\}}\ \ \underd{\{A/q^{j}\}\ldots\{A/q^{2j}\}}\ \
\underd{\{A/q^{2j+2}\}\ldots \{A/q^{b+j+1}\}}}
\end{array} \right.
\nn\\ \nn \\
  \nn \\
\left\{ \begin{array}{c}
 \frac{\lambda_{1,1}^{2m}}
{\underd{\{Aq^{a+2}\} \ldots \{Aq^4\}}\  \ \{Aq^2\} \ \ \{A\}\ \ \{A/q^2\}\ \
\underd{\{A/q^4\}\ldots\{A/q^{b+2}\}}}
\nn \\ \nn \\
\ldots \nn \\ \nn \\
\Longrightarrow \ \
\frac{\lambda_{i,1}^{2m}}
{\underd{\{Aq^{a+i+1}\} \ldots \{Aq^{2i+2}\}}\  \
\underd{\{Aq^{2i}\}\ldots \{Aq^{i+1}\}} \ \ \{Aq^{i-1}\}\ \ \{A/q^2\}\ \
\underd{\{A/q^4\}\ldots\{A/q^{b+2}\}}}
\end{array} \right.
\ee
Boxes contain products $\prod_k\{Aq^k\}$ over $k$ with no gaps. They turn into unity when
the lower limit exceeds the upper one, like $i+1>2i$ for $i=0$ or $2i+2>a+i+1$ for $i=a$.

\bigskip

Generalization to arbitrary $\lambda_{i,j}$, appearing in the single-floor pyramids,
is now obvious.
More complicated is adjusting the combinatorial coefficients, which guarantee
cancelation of poles and matching with the unknot, figure eight and trefoil.
By trial and error
we get the conjectural answer for arbitrary single-floor pyramid:
\be
{\cal F}_{\tiny\boxed{a\ \ldots \ 0 \ \ldots \ -b}}^{(m)} \ \stackrel{?}{=} \
\Big(q^{\frac{a-b}{2}}A\Big)^{a+b+1}
\left(\ \frac{1}{\underd{\{Aq^a\}\ldots\{A/q^b\}}}
\ \  + \right.
\label{firstfloorconj}
\ee
\be
\!\!\!\!\!\!\!\!\!\!\!\!\!\!
\left.
+\ \sum_{i=0}^a \sum_{j=0}^b
\frac{
(-)^{i+j+1}\Big(q^{i-j}A\Big)^{2m\cdot (i+j+1)}\,
\frac{[a]!}{[a-i]!i!}\cdot \frac{[b]!}{[b-j]![j]!}\cdot \frac{[a+b+1]}{[i+j+1]}}
{\underd{\{Aq^{a+i+1}\}\ldots\{Aq^{2i+2}\}}\
\underd{\{Aq^{2i}\}\ldots \{Aq^{i+1}\}}\
\{Aq^{i-j}\} \
\underd{\{A/q^{j+1}\}\ldots \{A/q^{2j}\}} \
\underd{\{A/q^{2j+2}\}\ldots \{A/q^{b+j+1}\}}
}
\ \right)
\nn \\
\nn
\ee
which fits all the expectations.
The item in the first line contains $\lambda_\emptyset^{2m}=1$ and can be considered
as associated with the zeroth floor.
This item is directly read from the pyramid itself, the eigenvalue leaves it intact.

\subsection{Other floors}

A natural generalization of peculiar $A$-dependent factor in
front of the power of the eigenvalue
$\ \lambda_{\stackrel{\ldots}{\stackrel{i_f\ j_f}{\stackrel{\ldots}{i_1\ \ \ j_1}}}}\ $
in the multi-floor analogue of (\ref{firstfloorconj}) is
\be
\frac{\{Aq^{2i+1}\}\{A/q^{2j+1}\}}{\{Aq^{i-j}\}} \ \ \longrightarrow \ \
\underline{
\prod_{f'\leq f''}  \frac{\{Aq^{i_{f'}+i_{f''}+1}\}\{A/q^{j_{f'}+j_{f''}+1}\}}
{\{Aq^{i_{f'}-j_{f''}}\}\{Aq^{i_{f''}-j_{f'}}\}}
}
\cdot \prod_f \frac{\{Aq^{2i_f+1}\}\{A/q^{2j_f+1}\}}{\{Aq^{i_f-j_f}\}}
\label{genfactor}
\ee
This gives the underlined "inter-floor interaction" factors like
\be
\lambda_{\stackrel{0,0}{1\   \ 1}}  & \longrightarrow & \frac{D_2D_{-2}}{D_1D_{-1}} \nn\\
\lambda_{\stackrel{0,0}{2\   \ 1}}  & \longrightarrow &\frac{D_3D_{-2}}{D_2D_{-1}} \nn\\
\lambda_{\stackrel{1,0}{2\   \ 1}}  & \longrightarrow &\frac{D_4D_{-2}}{D_2D_{0}} \nn\\
\ldots
\ee
Note, that they all preserve parity of powers in $q$.
Moreover, in empty cases, when $i_{f''}+j_{f''}+1=0$ the
underlined interaction factor drops away, because of the
cancelations between the numerator and denominator of (\ref{genfactor}).

\bigskip

Unfortunately, this is not exactly what needed  -- even if a floor in the eigenvalue
is empty, it still affects the shape of the contribution.
This is already seen in the fully { reliable} (not just conjectured) formulas (\ref{F22b}):
the non-trivial second-floor eigenvalue
$\lambda_{04}=\lambda_{\stackrel{0\ 0}{1\   \ 1}}$ appears only in the last term of
${\cal F}\!\!\!\!\!_{_{\tiny {\boxed{0\ }}\over{\boxed{1 \ 0\  -1}}}}$\!\!\!\!,
but the two other terms \\

\vspace{0.1cm}

\noindent
-- with the pure-first-floor eigenvalues $\lambda_2=\lambda_{\,1\ 0}$ and $\lambda_{-2}=\lambda_{\,0\ 1}$ --
are also not just the same as in ${\cal F}\,_{_{\tiny \boxed{1\ 0\ -1}}}$.

\bigskip

Another way to observe that the empty floor matters, is to compare the first two terms in
\be
{\cal F}_{_{\tiny\boxed{2\ 1\ 0 \ -1}}} = q^2A^4\cdot\left(\frac{1}{\{Aq^2\}\{Aq\}\{A\}\{A/q\}}
- \frac{[2]^2 \cdot \lambda_{0,0}^{2m}}{\{Aq^3\}\{Aq^2\}\{A\}\{A/q^2\}} + \ldots\right)
\label{F2101ex}
\ee
and the naive
\be
{\cal F}\!\!\!\!\!_{_{\tiny{\boxed{0}}\over{\boxed{2\ 1\ 0 \ -1}}}} \ \stackrel{???}{\approx}\
q^2 A^5\cdot\left(\frac{1}{\{Aq^2\}\{Aq\}\{A\}^2\{A/q\}}
- \frac{c\cdot \lambda_{0,0}^{2m}}{\{Aq^3\}\{Aq^2\}\{A\}^2\{A/q^2\}} + \ldots\right)
\label{F21010ex}
\ee
The coefficient $c$ in (\ref{F21010ex}) should be a quantization of $5$, but it should contain {\it odd}
powers of $q$, thus it can not be $[5]$.
In (\ref{F2101ex}) its analogue is a quantization of $4$, which is made out of {\it even} powers --
and it is $[2]^2$.
However, there is no such simple way out for $c$ in (\ref{F21010ex})
-- and already the second term in this formula is a less-naive combination of inverse differentials.
Formula (\ref{21010}) is an example.
In general, when one proceeds from ${\cal F}_{_{\tiny\boxed{a\ldots 0 \ldots -b}}}^{(m)}$ to\\

\vspace{0.25cm}

\noindent
${\cal F}^{(m)}_{_{\!\!\!\!\!\!\!\!\!\!\!\!\!\!\!\!\tiny{\boxed{0}}\over{
\!\!\!\!\!\!\!\!\boxed{a\ldots 0 \ldots -b}}}}$
the coefficient in front of
$\lambda_{00}^{2m}$ changes from $[a+b+1] $ to a sum of two items:
$[a+b+1]+\frac{D_aD_{-b}}{D_1D_{-1}}$
and to \ $[a+b+1] +[a_2+b_2+1]\frac{D_{a_1}D_{-b_1}}{D_{a_2+1}D_{-b_2-1}}$ \
for generic 2-floor pyramid.

\bigskip

This demonstrates that (\ref{genfactor}) is not the whole story in the multi-floor situation,
that additional guesses are needed -- and in fact easy to make, see Appendix B below.
Still, we prefer to postpone further speculations about the higher-floor contributions,
waiting for independent examination of the sequence of conjectures, already presented in this paper.

\newpage
\section{Conclusion}

To conclude, in this paper we made a new series of conjectures, which hopefully lead
to explicit formulas for exclusive Racah matrices $S$ and $\bar S$ and thus to construction
of arborescent knot polynomials in arbitrary rectangular representations.
This is a long standing problem, and its solution seems now within reach.
After the structure of differential expansion for defect-zero knots (\ref{41rs}) is revealed in this case
in \cite{rect41} and after the discovery of further factorization (\ref{factprop}) of its coefficients
for double braids in the present paper, it remains to conjecture these coefficients
for just a relatively simple family of twist knots.
To demonstrate that this can actually be possible,
we provided explicit expressions for {\it all} the contributions (\ref{F22a})-(\ref{210110})
in the case of
the previously unknown representation $R=[33]$
and made a general conjecture (\ref{firstfloorconj})
for the contributions for single-floor pyramids.
Its further extension to the second floor in Appendix B below can provide answers for arbitrary
arborescent knots in representations $R=[rr]$ and $R=[2^r]$.
Generic rectangular $R=[r^s]$ requires ${\rm min}(r,s)$ floors,
which will come as direct generalization.
Hopefully, this last step will be made soon enough.

\section*{Appendix A: Exclusive Racah matrix $\bar S^{[3,3]}$}

The Young diagram $R=[33]$ is not symmetric and there is no additional symmetry in Racah matrix:
it is just symmetric $\bar S_{ij}=\bar S_{ji}$.
The transformation $\bar S_{ij} \longrightarrow \pm \bar S_{11-j,11-i}$ converts it into something else.
%can now convert it into a matrix for $R=[222]$.

Standing in the first line/column are the square roots of quantum dimensions of the ten irreducible
representations, which constitute the product $[33]\otimes \overline{[33]}$, of which only six
were present in the case of $[22]\otimes \overline{[22]}$:

%\vspace{-0.5cm}
{\footnotesize
\be
\nn \\
d_{[33]}\bar S_{11} = \sqrt{\bar d_\emptyset} & & d_\emptyset =1 \nn \\
d_{[33]}\bar S_{21} = \sqrt{\bar d_{00}} & & d_{00}=   D_1D_{-1} \nn\\
d_{[33]}\bar S_{31} = \sqrt{\bar d_{10}} & & \bar d_{10} =\frac{D_{3}D_{0}^2D_{-1}}{[2]^2}\nn\\
d_{[33]}\bar S_{41} = \sqrt{\bar d_{01}} && \bar d_{01} =\frac{D_{1}D_{0}^2D_{-3}}{[2]^2} \nn \\
d_{[33]}\bar S_{51} = \sqrt{\bar d_{11}}& & \bar d_{11} =\frac{D_{3}D_{1}^2D_{-1}^2D_{-3}}{[3]^2} \nn \\
d_{[33]}\bar S_{61} = \sqrt{\bar d_{20}} && \bar d_{20} = \frac{D_5D_1^2D_0^2D_{-1}}{[3]^2[2]^2} \nn \\
d_{[33]}\bar S_{71} = \sqrt{\bar d_{\stackrel{00}{1\ 1}}}
&&\bar d_{\stackrel{00}{1\ 1}} =\frac{D_3D_{2}^2D_1D_{-1}D_{-2}^2D_{-3}}{[3]^2[2]^4 }\nn \\
d_{[33]}\bar S_{81} = \sqrt{\bar d_{21}}  && \bar d_{21} = \frac{D_5D_2^2D_0^2D_{-1}^2D_{-3}}{[4]^2[2]^2} \nn \\
d_{[33]}\bar S_{91} = \sqrt{\bar d_{\stackrel{00}{2\ 1}}}
&& \bar d_{\stackrel{00}{2\ 1}} = \frac{D_5D_3^2D_1D_0^2D_{-1}D_{-2}^2D_{-3}}{[4]^2[3]^2[2]^2} \nn \\
d_{[33]}\bar S_{10,1} = \sqrt{\bar d_{\stackrel{10}{2\ 1}}}
&\ \ \ \ \ \ \ \ \ \ \ \ \ \ \ \ \ \ \ \ \
& \bar d_{\stackrel{10}{2\ 1}} = \frac{D_5D_4^2D_3D_0^2D_{-1}^3D_{-2}^2D_{-3}}{[4]^2[3]^4[2]^4}
\nn \\ \nn
\ee
}

%\bigskip

\noindent
Most other entries are not fully factorizable and non-factorized pieces
can be expressed through $D_i=\{Aq^i\}/\{q\}$ in different ways.
The choices below are economic (with the exception of $\bar S_{55}$), but  {\it not canonical}:

{\footnotesize
\be
\nn \\
d_{[33]}\bar S_{12} = d_{[33]}\bar S_{21} \nn \\
d_{[33]}\bar S_{22} = \frac{D_1D_{-1}}{[3][2]D_3D_{-2}}\cdot\Big(([4]D_1+D_2)D_0-[3]^2[2]^2\Big)\nn \\
d_{[33]}\bar S_{32} =\frac{\sqrt{D_1D_0D_{-1}}}{[3][2]D_2D_{-2}\sqrt{D_3}}
\cdot\left(\frac{[4]}{[2]}D_1D_0-[3]^2[2]\right)\nn\\
d_{[33]}\bar S_{42} = \frac{D_1D_0\sqrt{D_{-3}D_{-1}}}{D_3D_{-2}}\cdot\Big(D_2D_0-[3]^2\Big)\nn\\
d_{[33]}\bar S_{52} = \frac{D_1D_{-1}\sqrt{D_1D_{-1}D_{-3}}}{[3][2]D_{-2}
\sqrt{D_3}}\cdot\Big(D_2D_0-[3][2]^2\Big)
\nn
\ee
\be
d_{[33]}\bar S_{62} = \frac{D_1D_0D_{-1}\sqrt{D_5D_1}}{[3][2]^2D_3D_{-2}}\cdot\Big(D_0^2-[3][2]^2\Big)  \nn \\
d_{[33]}\bar S_{72} = \frac{D_2D_1D_{-1}\sqrt{D_{-3}}}{[3]^2[2]^2\sqrt{D_3}}\cdot\Big(D_3D_0-[3]^2[2]\Big)  \nn \\
d_{[33]}\bar S_{82} = \frac{D_2D_0D_{-1}\sqrt{D_5D_1D_{-1}D_{-3}}}{[4][3][2]D_3D_{-2}}\cdot\Big(D_1D_0-[3]^2[2]\Big)  \nn \\
d_{[33]}\bar S_{92} = \frac{D_2D_1D_0D_{-1}\sqrt{D_5D_{-3}}}{[4][3]^2[2]^2D_2}\cdot\Big(D_2D_0-[3]^2[2]^2\Big)  \nn \\
d_{[33]}\bar S_{10,2} = -\frac{D_4D_0D_{-1}^2\sqrt{D_5D_1D_{-3}}}{[4][3][2]\sqrt{D_3}}
\nn \\ \nn
\ee
}
\vspace{-0.0cm}
{\footnotesize
\be
\nn \\
d_{[33]}\bar S_{13} = d_{[33]}\bar S_{31} \nn \\
d_{[33]}\bar S_{23} = d_{[33]}\bar S_{32} \nn \\
d_{[33]}\bar S_{33} =\frac{D_0^2 }{[3]^2[2]D_4D_{-2} }
\cdot\left( \frac{[4]}{[2]}D_4D_3D_{-1}D_{-3}-[4][2]D_4D_{-1}+[3]^2[2]  \right)\nn\\
d_{[33]}\bar S_{43} =\frac{D_0^2\sqrt{D_1D_{-1}D_{-3}} }{[3][2]^2D_{-2}\sqrt{D_3} }
\cdot\Big( D_3D_0-[4][2]^2  \Big)\nn\\
d_{[33]}\bar S_{53} = \frac{D_1D_0\sqrt{D_{-1}D_{-3}} }{[3]^2[2]^2D_4D_{-2} }
\cdot\Big( D_5D_1D_0D_{-1}-[3]^2([2]D_4+D_3)D_0+[4][3]^2[2]^2  \Big)\nn\\
d_{[33]}\bar S_{63} = \frac{D_1D_0^2\sqrt{D_5} }{ [3]^2[2]^2D_4D_{-2}\sqrt{D_3}}
\cdot\Big( D_4D_1D_0D_{-3}-[2]^3D_1D_0+[3]^3[2]^2   \Big)\nn\\
d_{[33]}\bar S_{73} = -\frac{D_2D_0\sqrt{D_1D_{-3}} }{[3][2]D_4 }\cdot\Big(D_4D_0-\frac{[4][3]}{[2]}\Big)\nn\\
d_{[33]}\bar S_{83} = \frac{D_2D_0^2\sqrt{D_5D_{-1}D_{-3}} }{ [4][3]^2[2]^2D_4D_{-2}\sqrt{D_3}}
\cdot\Big( D_3D_0-[4][2]^2  \Big)\Big(D_1D_0-[3]^2[2]\Big)\nn\\
d_{[33]}\bar S_{93} = -\frac{D_0^2\sqrt{D_5D_3D_1D_{-3}} }{[4][3][2]D_4 }\cdot\Big(D_3D_0-[4][2]^2 \Big)\nn\\
d_{[33]}\bar S_{10,3} =\frac{ D_0^2D_{-1}\sqrt{D_5D_{-3}}}{[4][2] } \nn\\
\nn \\ \nn
\ee
}
\vspace{-0.8cm}
{\footnotesize
\be
\nn \\
d_{[33]}\bar S_{14} = d_{[33]}\bar S_{41} \nn \\
d_{[33]}\bar S_{24} = d_{[33]}\bar S_{42} \nn \\
d_{[33]}\bar S_{34} = d_{[33]}\bar S_{43}\nn\\
d_{[33]}\bar S_{44} =\frac{D_1D_0^2 }{[4][3][2]D_3D_2D_{-2} }
\cdot\Big( [3]D_4D_3D_{-2}D_{-3} -[2]D_2D_{-1}+[4][3][2]  \Big)\nn\\
d_{[33]}\bar S_{54} = \frac{D_1D_0D_{-1}\sqrt{D_1} }{[4][3][2]D_2D_{-2}\sqrt{D_3} }
\cdot\Big( D_4D_2D_{-1}D_{-2} -[4]^2D_1D_0+[4]^2[2]^3   \Big)\nn\\
d_{[33]}\bar S_{64} = -\frac{[4]D_1D_0^2\sqrt{D_5D_1D_{-1}D_{-3}} }{[3][2]^2D_3D_{-2} } \nn\\
d_{[33]}\bar S_{74} = \frac{ D_1D_0\sqrt{D_3D_{-1}} }{[4][3]^2[2]^2D_3 }
\cdot\Big(D_4D_3D_{-1}D_{-2}-[4]^2D_2D_0+[4]^2[3][2]^2   \Big)\nn\\
d_{[33]}\bar S_{84} = -\frac{D_0^2D_{-1}\sqrt{D_5D_1} }{[3][2]D_3D_{-2} }\cdot\Big(D_1D_{-1}-[3]^2 \Big)\nn\\
d_{[33]}\bar S_{94} = - \frac{D_1D_0^2\sqrt{D_5D_{-1}} }{ [3]^2[2]^2D_2}\cdot\Big(D_2D_{-1}-[3]^2[2]\Big)\nn\\
d_{[33]}\bar S_{10,4} =\frac{D_4D_1D_0^2D_{-1}\sqrt{D_5D_{-1}} }{[3][2]^2D_2\sqrt{D_3D_1} } \nn\\
\nn \\ \nn
\ee
}

\vspace{-0.8cm}
{\footnotesize
\be
\nn \\
d_{[33]}\bar S_{15} = d_{[33]}\bar S_{51} \nn \\
d_{[33]}\bar S_{25} = d_{[33]}\bar S_{52} \nn \\
d_{[33]}\bar S_{35} = d_{[33]}\bar S_{53}\nn\\
d_{[33]}\bar S_{45} = d_{[33]}\bar S_{54} \nn\\
d_{[33]}\bar S_{55} =\frac{D_1^2D_{-1} }{[4][3]^2[2]^2D_4D_2D_{-2} }
\cdot\Big(D_5^2D_3D_{-3}^2D_{-1} - 3[3]^2D_5D_2D_0D_{-3}+[3]^2[2]^4D_3D_{-1} -\nn \\
\ \ \ \ \ \ - [3]^2\big(3q^8+10q^6+18q^4+23q^2+25+23q^{-2}+18q^{-4}+10q^{-6}+3q^{-8}\big)   \Big)\nn\\
d_{[33]}\bar S_{65} = -\frac{[4]D_1^2D_0 \sqrt{D_5D_{-1}D_{-3}}}{[3]^2[2]^2D_4D_{-2}\sqrt{D_3} }
\cdot\Big( D_2D_0-[3][2]^2  \Big)\nn\\
d_{[33]}\bar S_{75} = -\frac{D_1\sqrt{D_1D_{-1}} }{[4][3][2]D_4 }
\cdot\Big( D_4D_3D_0D_{-2}-[4]^2D_3D_0+[4]^2[3][2]  \Big)\nn\\
d_{[33]}\bar S_{85} = -\frac{D_1D_0D_{-1}\sqrt{D_5} }{[3]^2[2]^2D_4D_{-2}\sqrt{D_3} }
\cdot\Big( D_2D_1D_0^2-[3]^2[2]^2D_1D_0+[3]^3[2]^3  \Big)\nn\\
d_{[33]}\bar S_{95} =\frac{D_1D_0\sqrt{D_5D_3D_1D_{-1}} }{[3][2]D_4D_2 }\cdot\Big(D_2D_0-[3][2]^2\Big)\nn\\
d_{[33]}\bar S_{10,5} =-\frac{D_1D_0D_{-1}\sqrt{D_5D_{-1}} }{ [3]D_2}
\nn \\ \nn
\ee
}

\vspace{-0.7cm}
{\footnotesize
\be
d_{[33]}\bar S_{16} = d_{[33]}\bar S_{61} \nn \\
d_{[33]}\bar S_{26} = d_{[33]}\bar S_{62} \nn \\
d_{[33]}\bar S_{36} = d_{[33]}\bar S_{63}\nn\\
d_{[33]}\bar S_{46} = d_{[33]}\bar S_{64} \nn\\
d_{[33]}\bar S_{56} = d_{[33]}\bar S_{65}  \nn\\
d_{[33]}\bar S_{66} = \frac{D_1^2D_0 }{[4][3]^2[2]^2D_4D_3D_{-2} }
\cdot\Big( D_5D_4D_2D_0D_{-2}D_{-3}-[2]^2D_5D_4D_{-3}D_{-2}+[2]^2D_2D_0-[4][3][2]^3 \Big)\nn\\
d_{[33]}\bar S_{76} = \frac{[4]D_2D_1D_0\sqrt{D_5D_1D_{-3}} }{ [3][2]^2D_4\sqrt{D_3}} \nn\\
d_{[33]}\bar S_{86} = -\frac{D_2D_1D_0\sqrt{D_{-1}D_{-3}} }{[3]^2[2]^2D_4D_3D_{-2} }
\cdot\Big( D_5D_4D_{-1}D_{-2}-[2]D_3D_0+[4][2]^3  \Big)\nn\\
d_{[33]}\bar S_{96} =\frac{D_1D_0\sqrt{D_1D_{-3}} }{[3][2]^2D_4 }\cdot\Big( D_4D_0-[4][2]  \Big)\nn\\
d_{[33]}\bar S_{10,6} =-\frac{D_1D_0D_{-1}\sqrt{D_{-3}} }{[3][2]\sqrt{D_3} }  \nn
\ee
}
%\vspace{-0.5cm}

{\footnotesize
\be
d_{[33]}\bar S_{17} = d_{[33]}\bar S_{71} \nn \\
d_{[33]}\bar S_{27} = d_{[33]}\bar S_{72} \nn \\
d_{[33]}\bar S_{37} = d_{[33]}\bar S_{73}\nn\\
d_{[33]}\bar S_{47} = d_{[33]}\bar S_{74} \nn\\
d_{[33]}\bar S_{57} = d_{[33]}\bar S_{75}  \nn\\
d_{[33]}\bar S_{67} = d_{[33]}\bar S_{76} \nn\\
d_{[33]}\bar S_{77} = \frac{D_2D_1 }{[4][3][2]D_4D_3 }
\cdot\Big([3]D_5D_4D_{-1}D_{-2}-[2]D_3D_0+[4][3][2]\Big)\nn\\
d_{[33]}\bar S_{87} = \frac{D_2D_0\sqrt{D_5D_1D_{-1}} }{[3][2]^2D_4\sqrt{D_3} }
\cdot\Big( D_1D_0-[3]^2[2]\Big)\nn\\
d_{[33]}\bar S_{97} =-\frac{D_1D_0\sqrt{D_5} }{[3]D_4\sqrt{D_3} }\cdot\Big(D_2D_0-[3]^2   \Big)\nn\\
d_{[33]}\bar S_{10,7} =\frac{ D_0D_{-1}\sqrt{D_5D_1}}{[2]D_3 }   \nn
\ee
}
%\vspace{-0.5cm}
{\footnotesize
\be
d_{[33]}\bar S_{18} = d_{[33]}\bar S_{81} \nn \\
d_{[33]}\bar S_{28} = d_{[33]}\bar S_{82} \nn \\
d_{[33]}\bar S_{38} = d_{[33]}\bar S_{83}\nn\\
d_{[33]}\bar S_{48} = d_{[33]}\bar S_{84} \nn\\
d_{[33]}\bar S_{58} = d_{[33]}\bar S_{85}  \nn\\
d_{[33]}\bar S_{68} = d_{[33]}\bar S_{86} \nn\\
d_{[33]}\bar S_{78} = d_{[33]}\bar S_{87}  \nn\\
d_{[33]}\bar S_{88} = \frac{D_2D_0D_{-1} }{[3]^2[2]D_4D_3D_{-2} }
\cdot\left( \frac{[4]}{[2]}D_5D_3D_{-1}D_{-2}-[4][2]D_3D_{-2}+[3]^2[2]  \right)\nn\\
d_{[33]}\bar S_{98} =-\frac{D_0\sqrt{D_1D_{-1}} }{[3][2]D_4 }
\cdot\left(\frac{[4]}{[2]}D_2D_1-[3]^2[2]\right)\nn\\
d_{[33]}\bar S_{10,8} =\frac{D_0D_{-1}\sqrt{D_{-1}} }{[2]\sqrt{D_3} }   \nn
\ee
}
%\vspace{-0.5cm}

{\footnotesize
\be
%\nn \\
d_{[33]}\bar S_{19} = d_{[33]}\bar S_{91} \nn \\
d_{[33]}\bar S_{29} = d_{[33]}\bar S_{92} \nn \\
d_{[33]}\bar S_{39} = d_{[33]}\bar S_{93}\nn\\
d_{[33]}\bar S_{49} = d_{[33]}\bar S_{94} \nn\\
d_{[33]}\bar S_{59} = d_{[33]}\bar S_{95}  \nn\\
d_{[33]}\bar S_{69} = d_{[33]}\bar S_{96} \nn\\
d_{[33]}\bar S_{79} = d_{[33]}\bar S_{97}  \nn\\
d_{[33]}\bar S_{89} = d_{[33]}\bar S_{98}  \nn\\
d_{[33]}\bar S_{99} =\frac{D_1D_0 }{[3][2]D_4D_2 }\cdot\Big(\big([4]D_4+D_3\big)D_{-1}-[3]^2   \Big)\nn\\
d_{[33]}\bar S_{10,9} =-\frac{D_0D_{-1}\sqrt{D_1} }{D_2\sqrt{D_3} }\nn\\
\nn \\ \nn
\ee
}

\vspace{-1.0cm}
{\footnotesize
\be
\nn \\
d_{[33]}\bar S_{1,10} = d_{[33]}\bar S_{10,1} \nn \\
d_{[33]}\bar S_{2,10} = d_{[33]}\bar S_{10,2} \nn \\
d_{[33]}\bar S_{3,10} = d_{[33]}\bar S_{10,3}\nn\\
d_{[33]}\bar S_{4,10} = d_{[33]}\bar S_{10,4} \nn\\
d_{[33]}\bar S_{5,10} = d_{[33]}\bar S_{10,5}  \nn\\
d_{[33]}\bar S_{6,10} = d_{[33]}\bar S_{10,6} \nn\\
d_{[33]}\bar S_{7,10} = d_{[33]}\bar S_{10,7}  \nn\\
d_{[33]}\bar S_{8,10} = d_{[33]}\bar S_{10,8}  \nn\\
d_{[33]}\bar S_{9,10} = d_{[33]}\bar S_{10,9} \nn\\
d_{[33]}\bar S_{10,10} =\frac{D_0D_{-1} }{D_3D_2 }  \nn
\ee
}

%{\footnotesize$\gamma_1=[3]D_2D_{-2} - [2]^2, \ \gamma_2=D_2D_{-3} - [2], \ \gamma_3=D_3D_{-2} - [2],
% \ \gamma_4=D_3D_{-3} - 2[3] - 1, \ \gamma_5=D_3D_2D_{-2}D_{-3} - D_{3}D_{-3} + [2]^2$}.

%\vspace{-0.3cm}
\noindent
The complementary matrix is
\be
\bar T_{[3,3]} = {\rm diag}\Big(
 1, \ \ -A, \ \ q^2A^2, \ \ A^2/q^2, \ \ -A^3,\  \ -q^6A^3,\ \ A^4,\ \ q^4A^4, \ \ - q^4A^5,\ \ q^6A^6
\Big)
\ee
Then the eigenvalues of the product $\bar T^{[3,3]}\bar S^{[3,3]}\bar T^{[3,3]}$ in (\ref{SvsbS})  are:
\be
T^{-1}_{[3,3]} = A^{6} \cdot {\rm diag}\Big(
 q^{18}, \ \ -q^{14}, \ \ q^{12}, \ \ q^8, \ \ -q^6,\  \ q^2,\ \ -1,\ \ q^{-2}, \ \ - q^{-6},\ \ q^{-12}
\Big)
\ee
and the diagonalizing matrix $S^{[3,3]}$  can be obtained by the Cramer rule,
i.e. its entries are minors of the matrix $\bar T \bar S\bar T - T^{-1}$.
By Cramer rule  the matrix $M_{ij}$ with eigenvalues $\lambda_i$ is diagonalized
by a matrix $V_{jk} = {\rm Minor}_{jm}(M-\lambda_k\cdot I)$ with any $m$:
\be
\sum_j M_{ij}V_{jk} = \sum_j M_{ij}\cdot  {\rm Minor}_{jm}(M-\lambda_k\cdot I) \ \boxed{=}\
\sum_j \lambda_k \delta_{ij}{\rm Minor}_{jm}(M-\lambda_k\cdot I) = \lambda_k V_{ik}
\ee
where the boxed equality comes from
\be
\sum_j (M-\lambda_k\cdot I)_{ij}\cdot  {\rm Minor}_{jm}(M-\lambda_k\cdot I)
= \delta_{im} \det (M-\lambda_k\cdot I) = 0
\ee
and the r.h.s. vanishes because $\lambda_k$ is an eigenvalue.
In our case $M=\bar T \bar S\bar T$ and
\be
S_{ij} = \frac{{\rm Minor}_{\,im}(\,\bar T \,\bar S\,\bar T - T_j^{-1}\cdot I)}{\sigma_j^{(m)}}
\ee
The r.h.s. is actually independent of $m$ (modulo sign factors, depending on the definition of minors).
The matrix of minors is normalized by division over
$\sigma_j^{(m)} = \sqrt{\sum_i \Big({\rm Minor}_{\,im}(\,\bar T \,\bar S\,\bar T - T_j^{-1}\cdot I)\Big)^2}$
to  make $S$ orthogonal (but not symmetric).
Minors are determinants and they are relatively easy to calculate
(most important -- to simplify and factorize, what makes {\it this} diagonalization method most practical
in our situation).

With the matrix $S$ the first thing to calculate is
$H^{3_1}_{[3,3]}=d_{[3,3]} \cdot \Big(S_{[33]}T_{[33]}^{-3}S_{[33]}\Big)_{\emptyset\emptyset}$
for the trefoil.
In this way we reproduce the right answer,
which was the starting point for the entire consideration in \cite{rect41} --
thus closing the circle of reasoning, at least in the first previously unknown case of $R=[33]$.

\section*{Appendix B: Double-floor pyramids}

\subsubsection*{Maximal pyramids}

{\it Maximal} are the pyramids, where
all steps between the non-empty floors exactly equal to one, $a_{f+1}=a_f-1$ and $b_{f+1}=b_f-1$,
In this case the functions ${\cal F}_{\A,\B}$ possess additional symmetries.
For example in the case of maximal double-floor pyramids
${\cal F}\!_{_{\tiny{\boxed{{a-1}\ldots 0}\over\boxed{a\ \ \ \ldots \ -1}}}}$\!, \
which are relevant for description of representations $R=[rr]$,
this symmetry

\vspace{0.2cm}

\noindent
acts as $D_i\longrightarrow D_{2a-i-2}$ and identifies the coefficients in front of
\be
\lambda_{i0}^{2m} & \longleftrightarrow & \lambda_{\stackrel{a-i-2\ \ 0}{a\ \ \ \ \ \ \ \ \ 1}}^{2m} \nn\\
\lambda_{i1}^{2m} & \longleftrightarrow & \lambda_{\stackrel{a-i-2\ \ 0}{a-1\ \ \ \ \ \ 1}}^{2m} \nn \\
\ldots
\ee
Here the line $-1,0$ in the eigenvalue label is equivalent to $\emptyset$, for example
$\lambda_{-1,0}=\lambda_\emptyset=1$, $\lambda_{\stackrel{-1\ 0}{a\ \ \ \ 1}}=\lambda_{a1}$.
Note that the eigenvalues are {\it different}, equated by symmetry are the $A,q$-dependent {\it coefficients}
in front of them.

\bigskip

In order to respect this symmetry, the factors (\ref{genfactor}) should be complemented by additional
"small" corrections, which we put into boxes in formulas below.
Actually, the rule is simple:
\be
\lambda_{i0}^{2m} \ \longrightarrow \ \lambda_{i0}^{2m}\cdot\boxed{\frac{\{Aq^{i-1}\}}{\{A/q\}}}\ ,
\ \ \ \ \ \ \ \ \
\lambda_{i1}^{2m} \ \longrightarrow \
\lambda_{i1}^{2m}\cdot\boxed{\frac{\{Aq^i\}}{\{Aq^a\}}}\ ,
\ \ \ \ \ \ \ \ \
\lambda_{\stackrel{j0}{i\ 1}}^{2m} \ \longrightarrow \
\lambda_{\stackrel{j0}{i\ 1}}^{2m}\cdot\boxed{\frac{\{Aq^{i+j+1}\}}{\{Aq^{a+j+1}\}}}
\label{boxrule}
\ee
In accordance with this rule
\be
\!\!\!\!\!\!\!\!\!\!\!\!\!\!\!
{\cal F}_{_{\tiny\!\!\!\!\!\!{{\boxed{0}\ \ \ \ }\over{\boxed{1\ 0\ -1}}}}}^{\,(m)} =
 A^3\cdot \frac{A}{\{A\}}\cdot\left\{
 \left( \frac{1}{\{Aq\}\{A\}\{A/q\}}
  +\frac{\lambda_{\stackrel{00}{1\ 1}}^{2m}}{\{Aq^2\}\{A\}\{A/q^2\}}
  \cdot\frac{\{Aq^2\}\{A/q^2\}}{\{Aq\}\{A/q\}}
  \right) -
  \right.\nn \\ \left.
- [2]^2\left(\frac{  \lambda_{00}^{2m}}{\{Aq^2\}\{A\}\{A/q^2\}}
+ \frac{  \lambda_{11}^{2m}}{\{Aq^2\}\{A\}\{A/q^2\}}\right)
\right.
+ \nn \\  \nn \\ \left.
+ [3]\left(\frac{ \lambda_{10}^{2m} }{\{Aq^2\}\{Aq\} \{A/q^2\}}\cdot\boxed{\frac{\{A\}}{\{A/q\}}}
+ \frac{ \lambda_{01}^{2m} }{\{Aq^2\} \{A/q\}\{A/q^2\}}\cdot\boxed{\frac{\{A\}}{\{Aq\}}}
\right)
\right\}\nn
\ee

\be
{\cal F}_{\tiny\!\!\!\!{{\boxed{1\ 0}\ \ \ \ }\over{\boxed{2\ 1\ 0\ -1}}}}^{ (m)}
 \ \stackrel{?}{=} \
\frac{qA^2}{\{Aq\}\{A\}}\cdot q^2A^4\cdot \left\{
\left(
\frac{1}{\{Aq^2\}\{Aq\}\{A\}\{A/q\}}
+ \frac{\{A\}}{\{Aq^2\}}
\cdot\frac{\lambda_{\stackrel{10}{2\ 1}}^{2m}}{\{Aq^4\}\{Aq^3\} \{Aq\}\{A/q^2\} }
\cdot\frac{\{Aq^4\}\{A/q^2\}}{\{Aq^2\}\{A\}}\right)
- \right.\nn \\ \left.
- [3][2]\cdot\left(\frac{ \lambda_{00}^{2m}}{\{Aq^3\}\{Aq^2\} \{A\}\{A/q^2\}}
+  \frac{\{Aq\}}{\{Aq^2\}}\cdot
\frac{ \lambda_{\stackrel{00}{2\ 1}}^{2m}}{\{Aq^4\} \{Aq^3\} \{Aq\} \{A/q^2\}}
\cdot\frac{\{Aq^3\}\{A/q^2\}}{\{Aq^2\}\{A/q\}}\right)
+ \right.\nn \\ \left.
\!\!\!\!\!\!\!\!\!\!\!\!\!\!\!\!\!\!\!\!\!\!\!\!\!\!\!\!\!\!\!\!\!\!\!\!
+[3]^2\cdot \left( \frac{\lambda_{10}^{2m}  }{\{Aq^4\}\{Aq^2\}\{Aq\}\{A/q^2\}}
\cdot\boxed{\frac{\{A\}}{\{A/q\}}}
+ \frac{ \lambda_{21}^{2m}}{\{Aq^4\}\{Aq^3\}\{Aq\} \{A/q^2\}}\right)
+ \right.\nn \\ \left.
+\frac{[4][3]}{[2]}\cdot\left( \frac{ \lambda_{01}^{2m}  }{\{Aq^3\}\{Aq^2\}\{A/q\} \{A/q^2\}}
\cdot\boxed{ {\frac{\{A\}}{\{Aq^2\}}}}
+\frac{\{Aq\}}{\{Aq^2\}}\cdot
\frac{ \lambda_{\stackrel{00}{1\ 1}}^{2m}}{\{Aq^4\}\{Aq^2\} \{A\}\{A/q^2\}}
\cdot\frac{\{Aq^2\}\{A/q^2\}}{\{Aq\}\{A/q\}}
\cdot\boxed{ {\frac{\{Aq^2\}}{\{Aq^3\}}}}\right)
-\right.\nn
\ee
\vspace{-0.6cm}
\be
 \left.
- [4]\cdot \underline{\frac{ \lambda_{20}^{2m}}{\{Aq^4\}\{Aq^3\}\{Aq^2\} \{A/q^2\}}
\cdot\boxed{ {\frac{\{Aq\}}{\{A/q\}}}}}
\ \ \
-[4][2]^2\cdot \underline{\frac{ \lambda_{11}^{2m}}{\{Aq^4\}\{Aq^2\} \{A\}\{A/q^2\}}
\cdot\boxed{ {\frac{\{Aq\}}{\{Aq^2\}}}}}
\right\}
\label{210110a}
\ee

\be
\!\!\!\!\!\!\!\!\!\!\!\!\!\!\!\!\!\!
{\cal F}_{\tiny\!\!\!\!\!\!{{\boxed{2\ 1\ 0}\ \ }\over{\boxed{3\ 2\ 1\ 0\ -1}}}}^{(m)}  \ \stackrel{?}{=} \
\frac{q^3A^3}{\{Aq^2\}\{Aq\}\{A\}}\cdot
q^5A^5 \cdot \ \ \ \ \ \ \ \ \ \ \ \ \ \ \ \ \nn \\
\cdot\left\{
\left(\frac{1}{\{Aq^3\}\{Aq^2\}\{Aq\}\{A\}\{A/q\}}
+ \frac{\{Aq\}\{A\}}{\{Aq^4\}\{Aq^3\}}\cdot
\frac{\lambda_{\stackrel{20}{3\ 1}}^{2m}}{\{Aq^6\}\{Aq^5\}\{Aq^4\}\{Aq^2\}\{A/q^2\}}
\cdot\frac{\{Aq^6\}\{A/q^2\}}{\{Aq^3\}\{Aq\}}\right)
- \right. \nn \\
- [4][2]\cdot\left(\frac{\lambda_{00}^{2m}}{\{Aq^4\}\{Aq^3\}\{Aq^2\}\{A\}\{A/q^2\}}
+ \frac{\{A\}}{\{Aq^4\}}\cdot
\frac{\lambda_{\stackrel{10}{3\ 1}}^{2m}}{\{Aq^6\}\{Aq^5\}\{Aq^4\}\{Aq^2\}\{A/q^2\}}
\cdot\frac{\{Aq^5\}\{A/q^2\}}{\{Aq^3\}\{A\}}\right)
+ \nn \\
+\frac{[4][3]^2}{[2]}\cdot\left(\frac{\lambda_{10}^{2m}}{\{Aq^5\}\{Aq^4\}\{Aq^2\}\{Aq\}\{A/q^2\}}
\cdot\boxed{\frac{\{A\}}{\{A/q\}}}
+ \frac{\{Aq\}}{\{Aq^3\}}\cdot
\frac{\lambda_{\stackrel{00}{3\ 1}}^{2m}}{\{Aq^6\}\{Aq^5\}\{Aq^4\}\{Aq^2\}\{A/q^2\}}
\cdot\frac{\{Aq^4\}\{A/q^2\}}{\{Aq^3\}\{A/q\}}\right)
+ \nn \\
\!\!\!\!\!\!\!\!\!\!\!\!\!\!\!\!\!\!\!\!\!
+\frac{[5][4]}{[2]}\cdot\left(\frac{\lambda_{01}^{2m}}{\{Aq^4\}\{Aq^3\}\{Aq^2\}\{A/q\}\{A/q^2\}}
\cdot\boxed{\frac{\{A\}}{\{Aq^3\}}}
+\frac{\{A\}}{\{Aq^4\}}\cdot
\frac{\lambda_{\stackrel{10}{2\ 1}}^{2m}}{\{Aq^6\}\{Aq^4\}\{Aq^3\}\{Aq\}\{A/q^2\}}
\cdot\frac{\{Aq^4\}\{A/q^2\}}{\{Aq^2\}\{A\}}
\cdot\boxed{\frac{\{Aq^4\}}{\{Aq^5\}}}\right)
- \nn \\
\!\!\!\!\!\!\!\!\!\!\!\!\!\!\!\!\!\!\!\!\!\!\!\!\!\!\!\!\!\!\!\!\!\!\!\!\!\!\!\!\!\!\!\!\!\!\!\!\!\!
- [5][4][2]\left(\frac{\lambda_{11}^{2m}}{\{Aq^5\}\{Aq^4\}\{Aq^2\}\{A\}\{A/q^2\}}
\cdot\boxed{ \frac{\{Aq\}}{\{Aq^3\}} }
+ \frac{\{Aq\}}{\{Aq^3\}}\cdot
\frac{\lambda_{\stackrel{00}{2\ 1}}^{2m}}{\{Aq^6\}\{Aq^4\}\{Aq^3\}\{Aq\}\{A/q^2\}}
\cdot\frac{\{Aq^3\}\{A/q^2\}}{\{Aq^2\}\{A/q\}}
\cdot\boxed{\frac{\{Aq^3\}}{\{Aq^4\}} }
\right)
- \nn \\
\!\!\!\!\!\!\!\!\!\!\!\!\!\!\!\!\!\!\!\!\!\!\!\!\!
-[4]^2\cdot\left( \frac{\lambda_{20}^{2m}}{\{Aq^6\}\{Aq^4\}\{Aq^3\}\{Aq^2\}\{A/q^2\}}
\cdot\boxed{ \frac{\{Aq\}}{\{A/q\}} }
+ \frac{\lambda_{31}^{2m}}{\{Aq^6\}\{Aq^5\}\{Aq^4\}\{Aq^2\}\{A/q^2\}}\right)
+ \nn \\
\!\!\!\!\!\!\!\!\!\!\!\!\!\!\!\!\!\!\!\!\!\!\!\!\!\!\!\!\!\!\!\!\!\!
+ [5]\cdot\underline{ \frac{\lambda_{30}^{2m}}{\{Aq^6\}\{Aq^5\}\{Aq^4\}\{Aq^3\}\{A/q^2\}}
\cdot\boxed{\frac{\{Aq^2\}}{\{A/q\}} }}\ \ \
+ [5][3]^2\cdot\underline{ \frac{ \lambda_{21}^{2m}}{\{Aq^6\}\{Aq^4\}\{Aq^3\}\{Aq\}\{A/q^2\}}
\cdot\boxed{ \frac{\{Aq^2\}}{\{Aq^3\}} }} \ +
\nn
\ee
\vspace{-0.3cm}
\be
  \left.
+\ \frac{[5][4]^2}{[2]^2}\cdot\underline{\frac{\{Aq\}}{\{Aq^3\}}\cdot
\frac{\lambda_{\stackrel{00}{1\ 1}}^{2m}}{\{Aq^5\}\{Aq^4\}\{Aq^2\}\{A\}\{A/q^2\}}
\cdot\frac{\{Aq^2\}\{A/q^2\}}{\{Aq\}\{A/q\}}
\cdot\boxed{ \frac{ \{Aq^2\}}{ \{Aq^4\}} }}
\right\}
\label{32101210}
\ee

\bigskip

\noindent
The first two formulas already appeared in the main text -- in eqs.(\ref{F22b})
and (\ref{210110}) respectively.
Now we explicitly marked the extra correction factors, by putting them into boxes, and also
put the items related by the symmetry $D_i\longrightarrow D_{2a-i-2}$
into the same lines.
Self-symmetric items are underlined, they enter with their own coefficients.
The third formula is new, it contributes starting from $R=[44]$.

As to combinatorial coefficients, in this respect the function ${\cal F}$ for maximal pyramid is
\be
{\cal F}\!_{_{\tiny{\boxed{a-1\ldots 0}\over\boxed{a\ \ \ \ldots \ -1}}}} =
1 \ \oplus\ [2][a+1]\lambda_{00} \ \oplus \
\underbrace{\frac{[3][a+1][a]}{[2]}\lambda_{10} \ \oplus \ \frac{[a+2][a+1]}{[2]}\lambda_{10}} \oplus \nn\\
\ \oplus \ \underbrace{\frac{[2][a+2][a+1][a]}{[3]} \lambda_{11} \ \oplus \
\frac{[4][a+1][a][a-1]}{[2][3]}\lambda_{20}} \ \oplus \ \ldots
% \oplus \ ??? \lambda_{21} \ \oplus \ \ldots  \
%\oplus \ ??? \lambda_{\stackrel{00}{1\ 1}} \ \oplus \ ??? \lambda_{\stackrel{00}{2\ 1}}
%\ \oplus \ ??? \lambda_{\stackrel{10}{2\ 1}} \ \oplus \ \ldots
\ee
where suppressed are the powers $2m$ and the combinations of differentials.
The sums of underbraced coefficients at $q=1$ are binomial coefficients $C^k_{2a+2}$
for $k=0,1,2,3,\ldots$,
as usual for the differential expansions.

\subsection*{Minimal pyramids}

Needed for complete description of the case $R=[44]$  are two more functions with no obvious symmetry.
One of them is associated with the {\it minimal} pyramid, where the second floor has just one box:
\be
{\cal F}_{\tiny\!\!\!\!\!\!{{\ \ \ \boxed{0}\ \ }\over{\boxed{3\ 2\ 1\ 0\ -1}}}}^{(m)}
  \ \stackrel{?}{=} \
\frac{A}{\{A\}}\cdot
q^5A^5\cdot\left(
\frac{1}{\{Aq^3\}\{Aq^2\}\{Aq\}\{A\}\{A/q\}}
- \frac{\lambda_{00}^{2m}}{\{Aq^4\}\{Aq^3\}\{Aq^2\}\{A\}\{A/q^2\}}
\cdot\boxed{\boxed{\left([5]+\frac{\{Aq^3\}}{\{Aq\}}\right)}}\
+ \right.\nn \\
+\frac{\frac{[5][4]}{[2]}\lambda_{10}^{2m}}{\{Aq^5\}\{Aq^4\}\{Aq^2\}\{Aq\}\{A/q^2\}}
\cdot \boxed{\frac{\{A\}}{\{A/q\}}}
+\frac{[5]\,\lambda_{01}^{2m}}{\{Aq^4\}\{Aq^3\}\{Aq^2\}\{A/q\}\{A/q^2\}}
\cdot \boxed{\frac{\{A\}}{\{Aq\}}}\
- \nn \\
- \frac{\frac{[5][4]}{[3]} \lambda_{20}^{2m}}{\{Aq^6\}\{Aq^4\}\{Aq^3\}\{Aq^2\}\{A/q^2\}}
\cdot \boxed{\frac{\{A\}}{\{A/q\}}}\
-\frac{\frac{[5][4][2]}{[3]}\lambda_{11}^{2m}}{\{Aq^5\}\{Aq^4\}\{Aq^2\}\{A\}\{A/q^2\}}
+ \nn \\
%\!\!\!\!\!\!\!\!\!\!\!\!\!\!\!\!\!\!\!\!\!\!\!\!
+ \frac{\frac{[5]}{[3]}\lambda_{30}^{2m}}{\{Aq^6\}\{Aq^5\}\{Aq^4\}\{Aq^3\}\{A/q^2\}}
\cdot \boxed{\frac{\{A\}}{\{A/q\}}}\
+ \frac{[5][2]\,\lambda_{21}^{2m}}{\{Aq^6\}\{Aq^4\}\{Aq^3\}\{Aq\}\{A/q^2\}}
+ \nn \\
+ \frac{\frac{[5][4]}{[3][2]}\lambda_{\stackrel{00}{1\ 1}}^{2m}}{\{Aq^5\}\{Aq^4\}\{Aq^2\}\{A\}\{A/q^2\}}
\cdot\frac{\{Aq^2\}\{A/q^2\}}{\{Aq\}\{A/q\}}
- \nn \\
-\frac{\frac{[4][2]}{[3]}\lambda_{31}^{2m}}{\{Aq^6\}\{Aq^5\}\{Aq^4\}\{Aq^2\}\{A/q^2\}}
- \frac{\frac{[5][2]}{[3]}\lambda_{\stackrel{00}{2\ 1}}^{2m}}{\{Aq^6\}\{Aq^4\}\{Aq^3\}\{Aq\}\{A/q^2\}}
\cdot\frac{\{Aq^3\}\{A/q^2\}}{\{Aq^2\}\{A/q\}} +
\nn
\ee
\vspace{-0.4cm}
\be
\ \ \ \ \ \ \ \ \ \ \ \ \ \ \ \ \ \
 \left.
+\ \frac{\lambda_{\stackrel{00}{3\ 1}}^{2m}}{\{Aq^6\}\{Aq^5\}\{Aq^4\}\{Aq^2\}\{A/q^2\}}
\cdot\frac{\{Aq^4\}\{A/q^2\}}{\{Aq^3\}\{A/q\}}
\right)
\label{321010}
\ee

\noindent
Double-boxed  are correction factors, deviating from the rule (\ref{boxrule}).
To reveal their shape in the case of minimal pyramids
we present in the same form (with explicitly shown correction terms) the function (\ref{21010}):
\be
{\cal F}_{\tiny\!\!\!\!\!\!\!\!{{\boxed{0}\ \ }\over{\boxed{2\ 1\ 0\ -1}}}}^{(m)}  \ \stackrel{?}{=} \
\frac{A}{\{A\}}\cdot q^2A^4\cdot\left(
\frac{1}{\{Aq^2\}\{Aq\}\{A\}\{A/q\}}
- \frac{\,\lambda_{00}^{2m}}{\{Aq^3\}\{Aq^2\}\{A\}\{A/q^2\}}
\cdot\boxed{\boxed{\left([4]+\frac{\{Aq^2\}}{\{Aq\}}\right)}}\
%- \frac{ \lambda_{00}^{2m}}{\{Aq^3\}\{Aq^2\}\{A\}\{A/q^2\}}\boxed{\frac{\{Aq^2\}}{\{Aq\}}}
+ \right.\nn \\ \left.
\!\!\!\!\!\!\!\!\!\!\!\!\!\!\!\!\!\!\!\!\!\!\!\!\!\!\!\!\!\!\!\!\!\!\!\!
+ \ \frac{\frac{[4][3]}{[2]} \lambda_{10}^{2m}  }{\{Aq^4\}\{Aq^2\}\{Aq\} \{A/q^2\}}
\cdot\boxed{\frac{\{A\}}{\{A/q\}}}\
+ \frac{[4] \lambda_{01}^{2m}  }{\{Aq^3\}\{Aq^2\} \{A/q\}\{A/q^2\}}
\cdot\boxed{\frac{\{A\}}{\{Aq\}}} \
- \right.\nn \\ \left.
- \ \frac{\frac{[4]}{[2]} \lambda_{20}^{2m}}{\{Aq^4\}\{Aq^3\}\{Aq^2\} \{A/q^2\}}
\cdot\boxed{\frac{\{A\}}{\{A/q\}}} \
-\frac{[4][2] \lambda_{11}^{2m}}{\{Aq^4\}\{Aq^2\}\{A\}\{A/q^2\}} \
+ \right.\nn \\
%\!\!\!\!\!\!\!\!\!\!\!\!\!\!\!\!\!\!\!\!\!\!\!\!
+ \ \frac{[3] \lambda_{21}^{2m}}{\{Aq^4\}\{Aq^3\}\{Aq\}\{A/q^2\}}
+  \frac{\frac{[4]}{[2]}\lambda_{\stackrel{00}{1\ 1}}^{2m}}{\{Aq^4\}\{Aq^2\} \{A\}\{A/q^2\}}
\cdot\frac{\{Aq^2\}\{A/q^2\}}{\{Aq\}\{A/q\}} - \nn \\ \left.
- \frac{\lambda_{\stackrel{00}{2\ 1}}^{2m}}{\{Aq^4\}\{Aq^3\}\{Aq\} \{A/q^2\}}
\cdot\frac{\{Aq^3\}\{A/q^2\}}{\{Aq^2\}\{A/q\}}
\right)
\label{21010a}
\ee

%From these examples, a plausible modification of (\ref{boxrule})
%for generic $a_2$, not obligatory equal to $a_1-1$, is
%\be
%\lambda_{i0}^{2m} \ \longrightarrow \
%\lambda_{i0}^{2m}\cdot\boxed{\frac{\{Aq^{i-1}\}}{\{A/q\}}}\ ,
%\ \ \ \ \ \ \ \ \
%\lambda_{i1}^{2m} \ \longrightarrow \
%\lambda_{i1}^{2m}\cdot\boxed{\frac{\{Aq^{i}\}}{\{Aq^{a_2+1}\}}}\
%\nn \\
%\ \ \ \ \ \ \ \ \
%\lambda_{\stackrel{i_2\ 0}{i_1\ \ 1}}^{2m} \ \longrightarrow \
%\lambda_{\stackrel{i_2\ 0}{i_1\ \ 1}}^{2m}\cdot\boxed{\frac{\{Aq^{i_1+i_2+1}\}}{\{Aq^{a_2+i_2+2}\}}}
%\label{boxrule2}
%\ee

\noindent
Now it is easy to guess the rule, which substitutes (\ref{boxrule}) for
minimal two-floor pyramids: if
\be
{\cal F}^{(m)}_{_{\tiny\boxed{a\ \ldots \ -1}}} = \alpha_\emptyset(A,q) +
\sum_{i=0}^a\sum_{j=0}^1 \alpha_{ij}(A,q) \cdot\lambda_{ij}^{2m}
\ee
with $\alpha_{ij}$ explicitly given in (\ref{firstfloorconj}), then
\be
{\cal F}^{(m)}_{_{\tiny{\boxed{0}\over{\!\!\!\!\boxed{a\ \ldots \ 0 \ -1}}}}} \ \stackrel{?}{=}\
\frac{A}{\{A\}}\cdot\left(
\alpha_\emptyset(A,q) +
\sum_{i=0}^a\sum_{j=0}^1 \alpha_{ij}^{^{\tiny\boxed{0}}}(A,q)\cdot \lambda_{ij}^{2m}
+ \sum_{i=1}^a \beta_{i1}^{\,^{\tiny\boxed{0}}}(A,q) \cdot
\big(\underbrace{\lambda_{00}\lambda_{i1}}_{\lambda_{\stackrel{00}{i\ 1}}}\big)^{2m}
\right)
\label{F0}
\ee
with
\vspace{-0.6cm}
\be
& \alpha^{^{\tiny\boxed{0}}}_{00} = \alpha_{00}
\cdot \boxed{\boxed{\left(1+\frac{1}{[a+2]}\frac{\{Aq^a\}}{\{Aq\}}\right)}} \nn \\
1\leq i \leq a:& \alpha^{^{\tiny\boxed{0}}}_{i0} = \alpha_{i0}
\cdot \boxed{\boxed{\frac{[a+1]}{[a]}}\cdot \frac{\{A\}}{\{A/q\}}} \nn \\
& \alpha^{^{\tiny\boxed{0}}}_{01} = \alpha_{01}\cdot \boxed{\boxed{[2]}\cdot\frac{\{A\}}{\{Aq\}}} \nn \\
1\leq i \leq a:& \alpha^{^{\tiny\boxed{0}}}_{i1} = \alpha_{i1}\cdot
\boxed{\boxed{\frac{[2][a+1]}{[a]}}} \nn \\
1\leq i \leq a:& \ \ \ \ \beta^{\,^{\tiny\boxed{0}}}_{i1} = \alpha_{i1}\cdot
\frac{\{Aq^{i+1}\}\{A/q^2\}}{\{Aq^i\}\{A/q\}}\cdot\boxed{\boxed{\frac{[a+1]}{[a]}\cdot\frac{[i]}{[i+1]}}}
\label{dboxmin}
\ee

%\bigskip

%\noindent
%At least at this stage additional correction factors in double boxes can be considered
%as a change of combinatorial coefficients.

\subsection*{Intermediate pyramid, contributing for $R=[44]$}

The last contribution, needed in the case of $R=[44]$, is
\be
\!\!\!\!\!\!\!\!\!\!\!\!\!\!\!\!
{\cal F}_{\tiny\!\!\!\!\!\!\!\!{{\boxed{1\ 0}\ \ }\over{\boxed{3\ 2\ 1\ 0\ -1}}}}^{(m)}  \ \stackrel{?}{=} \
\frac{qA^2}{\{Aq\}\{A\}}\cdot
q^5A^5\cdot\left(
\frac{1}{\{Aq^3\}\{Aq^2\}\{Aq\}\{A\}\{A/q\}}
- \frac{\lambda_{00}^{2m}}{\{Aq^4\}\{Aq^3\}\{Aq^2\}\{A\}\{A/q^2\}}
\cdot\boxed{\boxed{\left([5]+[2]\cdot\frac{\{Aq^3\}}{\{Aq^2\}}\right)}}\
+ \right.\nn \\
+ \ \frac{\frac{[3]}{[2]}\,\lambda_{10}^{2m}}{\{Aq^5\}\{Aq^4\}\{Aq^2\}\{Aq\}\{A/q^2\}}
\cdot\boxed{\frac{\{A\}}{\{A/q\}}\cdot\boxed{\left([6]+[3]\cdot\frac{\{Aq^3\}}{\{Aq^2\}} \right)  }}
+\frac{\frac{[5][3]}{[2]}\,\lambda_{01}^{2m}}{\{Aq^4\}\{Aq^3\}\{Aq^2\}\{A/q\}\{A/q^2\}}
\cdot\boxed{\frac{\{A\}}{\{Aq^2\}}}\
- \nn \\
- \ \frac{\frac{[5][4]}{[2]}\,\lambda_{20}^{2m}}{\{Aq^6\}\{Aq^4\}\{Aq^3\}\{Aq^2\}\{A/q^2\}}
\cdot\boxed{\frac{\{Aq\}}{\{A/q\}}}
- \ \frac{[5]\,\lambda_{11}^{2m}}{\{Aq^5\}\{Aq^4\}\{Aq^2\}\{A\}\{A/q^2\}}
\cdot\boxed{ \frac{\{Aq\}}{\{Aq^2\}}\cdot\boxed{\left([4]+\frac{\{Aq^4\}}{\{Aq^3\}}\right)}}\
+ \nn \\
%\!\!\!\!\!\!\!\!\!\!\!\!\!\!\!\!\!\!\!\!\!\!\!\!\!\!\!\!\!
+ \ \frac{\frac{[5]}{[2]}\,\lambda_{30}^{2m}}{\{Aq^6\}\{Aq^5\}\{Aq^4\}\{Aq^3\}\{A/q^2\}}
\cdot\boxed{\frac{\{Aq\}}{\{A/q\}}}\
+ \ \frac{\frac{[5][3]^2}{[2]}\,\lambda_{21}^{2m}}{\{Aq^6\}\{Aq^4\}\{Aq^3\}\{Aq\}\{A/q^2\}}
+ \nn \\
+ \ \frac{\{Aq\}}{\{Aq^2\}}\cdot
\frac{\frac{[5][4]}{[2]}\,\lambda_{\stackrel{00}{1\ 1}}^{2m}}{\{Aq^5\}\{Aq^4\}\{Aq^2\}\{A\}\{A/q^2\}}
\cdot\frac{\{Aq^2\}\{A/q^2\}}{\{Aq\}\{A/q\}}
\cdot\boxed{\frac{\{Aq^2\}}{\{Aq^3\}}}\
-\nn \\
- \ \frac{\frac{[4][3]}{[2]}\,\lambda_{31}^{2m}}{\{Aq^6\}\{Aq^5\}\{Aq^4\}\{Aq^2\}\{A/q^2\}}
- \ \frac{\{Aq\}}{\{Aq^2\}}\cdot
\frac{[5][3]\,\lambda_{\stackrel{00}{2\ 1}}^{2m}}{\{Aq^6\}\{Aq^4\}\{Aq^3\}\{Aq\}\{A/q^2\}}
\cdot\frac{\{Aq^3\}\{A/q^2\}}{\{Aq^2\}\{A/q\}}\
+ \nn \\
\!\!\!\!\!\!\!\!\!\!\!\!\!\!\!\!\!\!\!\!\!\!\!
+\ \frac{\{Aq\}}{\{Aq^2\}}\cdot
\frac{\frac{[3]^2}{[2]}\,\lambda_{\stackrel{00}{3\ 1}}^{2m}}{\{Aq^6\}\{Aq^5\}\{Aq^4\}\{Aq^2\}\{A/q^2\}}
\cdot\frac{\{Aq^4\}\{A/q^2\}}{\{Aq^3\}\{A/q\}}
+\frac{\{A\}}{\{Aq^2\}}\cdot
\frac{\frac{[5]}{[2]}\,\lambda_{\stackrel{10}{2\ 1}}^{2m}}{\{Aq^6\}\{Aq^5\}\{Aq^4\}\{Aq^2\}\{A/q^2\}}
\cdot\frac{\{Aq^4\}\{A/q^2\}}{\{Aq^2\}\{A\}}\ - \
\nn
\ee
\vspace{-0.4cm}
\be \left.
- \frac{\{A\}}{\{Aq^2\}}\cdot
\frac{\lambda_{\stackrel{10}{3\ 1}}^{2m}}{\{Aq^6\}\{Aq^5\}\{Aq^4\}\{Aq^2\}\{A/q^2\}}
\cdot\frac{\{Aq^5\}\{A/q^2\}}{\{Aq^3\}\{A\}}
\right)
\label{3210110}
\ee
This formula implies a simple adjustment  of the rule (\ref{boxrule}) to the case $a_2< a_1-1$:
\be
\lambda_{i0}^{2m} \ \longrightarrow \ \lambda_{i0}^{2m}
\cdot\boxed{\frac{\{Aq^{{\rm min}(i-1,a_2)}\}}{\{A/q\}}}\ ,
\ \ \ \ \ \ \
\lambda_{i1}^{2m} \ \longrightarrow \
\lambda_{i1}^{2m}\cdot\boxed{\frac{\{Aq^{{\rm min}(i,a_2+1)}\}}{\{Aq^{a_2+1}\}}}\ ,
\nn\\
\lambda_{\stackrel{i_2\ 0}{i_1\ \ 1}}^{2m} \ \longrightarrow \
\lambda_{\stackrel{i_2\ 0}{i_1\ \ 1}}^{2m}\cdot
\boxed{\frac{\{Aq^{{\rm min}(i_1+i_2+1,a_2+i_2+2)}\}}{\{Aq^{a_2+i_2+2}\}}}
\label{boxrule2}
\ee
The second formula follows from the third one, because the empty line in the eigenvalue label corresponds
to $i_2+j_2+1=0$ and  in our current examples $j_2=0$.

Also the next-level corrections in double boxes
are getting richer than in (\ref{dboxmin}) and need to be tamed,
what is not so difficult to do.
For example, the coefficient in front of $\lambda_{00}^{2m}$ is
\be
\phantom. [a_1+2] + [a_2+1]\frac{\{Aq^{a_1}\}}{\{Aq^{a_2+1}\}}
\label{dboxcor}
\ee
what interpolates between $[a_1+2]$ when the second flour is absent, $a_2+1=0$,
and $[a_1+2]+[a_1]=[2][a_1+1]$ for the maximal pyramid with $a_2=a_1-1$.

\subsection*{Formulas for arbitrary $a=a_1$}

The next step one can make is to fix the second floor, but release the first,
i.e. write a formula for generic $a=a_1$ at fixed $a_2$, like we already did in
(\ref{firstfloorconj}) for $a_2=-1$ (empty second floor) and (\ref{dboxmin}) for $a_2=0$
(one box at the second floor).

The analogue of (\ref{dboxmin}) for generic pyramids with two boxes ($a_2=1$)
at the second floor is
\be
{\cal F}^{(m)}_{_{\!\!\!\!\!\!\!\!\!\!\!\!\tiny{\boxed{1\ 0}\over{\!\!\!\!\boxed{a\ \ldots \ 1\ 0 \ -1}}}}}
\!\!\!\!\!\!
\stackrel{?}{=}\
\frac{qA^2}{\{Aq\}\{A\}}\cdot\left(
\alpha_\emptyset  +
\sum_{i=0}^a\sum_{j=0}^1 \alpha_{ij}^{^{\tiny\boxed{10}}} \cdot \lambda_{ij}^{2m}
+ \sum_{i=1}^a \beta_{i1}^{\,^{\tiny\boxed{10}}}  \cdot
\big(\underbrace{\lambda_{00}\lambda_{i1}}_{\lambda_{\stackrel{00}{i\ 1}}}\big)^{2m}
%+ \right. \nn \\ \left.
+ \sum_{i=2}^a \gamma_{i1}^{\,^{\tiny\boxed{10}}}  \cdot
\big(\underbrace{\lambda_{10}\lambda_{i1}}_{\lambda_{\stackrel{10}{i\ 1}}}\big)^{2m}
\right)
\label{F10}
\ee
with
\be
& \alpha^{^{\tiny\boxed{10}}}_{00}\ \stackrel{(\ref{dboxcor})}{ =}\  \alpha_{00}
\cdot \boxed{\boxed{\left(1+\frac{[2]}{[a+2]}\frac{\{Aq^a\}}{\{Aq^2\}}\right)}} \nn \\
& \!\!\!\!\!\!\!\!\!\!\!\!\!\!\!\! \alpha^{^{\tiny\boxed{10}}}_{10} = \alpha_{10}
\cdot \boxed{\boxed{\frac{1}{[a+2][a]}\cdot\frac{[3][a]\{Aq^3\}\{Aq^a\}
- [2][a-2]\{Aq^2\}\{Aq^a\} + [a+2][a+1]\{Aq^2\}\{Aq^3\}}{\{Aq^2\}\{Aq^3\}}}\cdot \frac{\{A\}}{\{A/q\}}} \nn \\
2\leq i \leq a:& \alpha^{^{\tiny\boxed{10}}}_{i0} = \alpha_{i0}
\cdot \boxed{\boxed{\frac{[a+1]}{[a-1]}}\cdot \frac{\{Aq\}}{\{A/q\}}} \nn \\
& \alpha^{^{\tiny\boxed{10}}}_{01} = \alpha_{01}\cdot \boxed{\boxed{[3]}\cdot\frac{\{A\}}{\{Aq^2\}}} \nn \\
 & \alpha^{^{\tiny\boxed{10}}}_{11} = \alpha_{11}\cdot
\boxed{\frac{\{Aq\}}{\{Aq^2\}}\cdot \boxed{\frac{[3][a+1]}{[a]}
\cdot\left(1+\frac{1}{[a+1]}\frac{\{Aq^{a+1}\}}{\{Aq^3\}}\right)}} \nn \\
2\leq i \leq a:& \alpha^{^{\tiny\boxed{10}}}_{i1} = \alpha_{i1}\cdot
\boxed{\boxed{\frac{[3][a+1]}{[a-1]}} }
\nn
\ee
\be
& \ \ \ \ \beta^{\,^{\tiny\boxed{10}}}_{11} =
\frac{\{Aq\}}{\{Aq^2\}}\cdot \alpha_{11}\cdot\frac{\{Aq^{2}\}\{A/q^2\}}{\{Aq\}\{A/q\}}
\cdot\boxed{\boxed{\frac{[3][a+1]}{[2][a]} }\cdot \frac{\{Aq^2\}}{\{Aq^3\}}}\nn\\
2\leq i \leq a:& \ \ \ \ \beta^{\,^{\tiny\boxed{10}}}_{i1} =
\frac{\{Aq\}}{\{Aq^2\}}\cdot \alpha_{i1}\cdot\frac{\{Aq^{i+1}\}\{A/q^2\}}{\{Aq^i\}\{A/q\}}
\cdot\boxed{\boxed{\frac{[3][a+1]}{[a-1]}\cdot\frac{[i]}{[i+1]}}}\nn\\
2\leq i \leq a:& \ \ \ \ \gamma^{\,^{\tiny\boxed{10}}}_{i1} =
\frac{\{A\}}{\{Aq^2\}}\cdot \alpha_{i1}\cdot\frac{\{Aq^{i+2}\}\{A/q^2\}}{\{Aq^i\}\{A\}}
\cdot\boxed{\boxed{\frac{[a+1]}{[a-1]}\cdot\frac{[i-1]}{[i+1]}}}
\label{dbox2box}
\ee
Like in the previous cases in this Appendix, we present the formula only for $b_1=1$,
since this is sufficient for the study of $R=[rr]$ representations.
Like in the case of Racah matrices in Appendix A, the formula for $\alpha_{10}$, which
is non-linear (this time -- quadratic) in the differentials $\{Aq^2\}$ and $\{Aq^3\}$
can be written in many different ways.
To find a canonical writing one needs to work out some more examples -- for bigger pyramids.
Emerging formulas can look lengthy and complicated, as compared to expressions, which we
encountered earlier, for particular small values of $a$, but in fact they are much
better structured.

\bigskip

Namely, for three boxes at the second floor we get
\be
{\cal F}^{(m)}_{_{\!\!\!\!\!\!\!\!\!\!\!\!\tiny{\boxed{2\ 1\ 0}\over{\!\!\!\!\boxed{a\ \ldots \ 2\ 1\ 0 \ -1}}}}}
\!\!\!\!\!\!
\stackrel{?}{=}\
\frac{q^3A^3}{\{Aq^2\}\{Aq\}\{A\}}\cdot\left(
\alpha_\emptyset  +
\sum_{i=0}^a\sum_{j=0}^1 \alpha_{ij}^{^{\tiny\boxed{210}}} \cdot \lambda_{ij}^{2m}
+ \sum_{i=1}^a \beta_{i1}^{\,^{\tiny\boxed{210}}}  \cdot
\big(\underbrace{\lambda_{00}\lambda_{i1}}_{\lambda_{\stackrel{00}{i\ 1}}}\big)^{2m}
+ \right. \nn \\ \left.
+ \sum_{i=2}^a \gamma_{i1}^{\,^{\tiny\boxed{210}}}  \cdot
\big(\underbrace{\lambda_{10}\lambda_{i1}}_{\lambda_{\stackrel{10}{i\ 1}}}\big)^{2m}
+ \sum_{i=3}^a \delta_{i1}^{\,^{\tiny\boxed{210}}}  \cdot
\big(\underbrace{\lambda_{20}\lambda_{i1}}_{\lambda_{\stackrel{20}{i\ 1}}}\big)^{2m}
\right)
\label{F210}
\ee
with
\be
& \alpha^{^{\tiny\boxed{210}}}_{00}\ \stackrel{(\ref{dboxcor})}{ =}\  \alpha_{00}
\cdot \boxed{\boxed{\left(1+\frac{[3]}{[a+2]}\frac{\{Aq^a\}}{\{Aq^3\}}\right)}} \nn \\
& \!\!\!\!\!\!\!\!\!\!\!\!\!\!\!\! \!\!\!\!\!\!\!\!\!\!\!\!\!\!\!\! \!\!\!\!\!\!\!\!\!\!\!\!\!\!\!\!
\!\!\!\!\!\!\!\!\!\!\!   \alpha^{^{\tiny\boxed{210}}}_{10} = \alpha_{10}
\cdot \boxed{\boxed{\frac{1}{[a+2][a]}\cdot\frac{[2][4][a-1]\{Aq^4\}\{Aq^a\}
- [2][3][a-3]\{Aq^3\}\{Aq^a\} + [a+2][a+1]\{Aq^4\}\{Aq^3\}}{\{Aq^4\}\{Aq^3\}}}\cdot \frac{\{A\}}{\{A/q\}}} \nn \\
& \!\!\!\!\!\!\!\!\!\!\!\!\!\!\!\! \!\!\!\!\!\!\!\!\!\!\!\!\!\!\!\! \!\!\!\!\!\!\!\!\!\!\!\!\!\!\!\!
 \!\!\!\!\!\!\!\!\!\!\!\!\!\!\!\!\!\!\!\!\!\!\!\alpha^{^{\tiny\boxed{210}}}_{20} = \alpha_{20}
\cdot \boxed{\boxed{\!\! \begin{array}{l}
\ \ \ \ \frac{1}{[a+2][a][a-1][a-2]}\cdot\Big([4][a+1][a][a-1][a-2]\{Aq^3\}\{Aq^4\}\{Aq^5\}
-\frac{[4][3]}{[2]}[a][a-1][a-2][a-3]\{Aq^2\}\{Aq^4\}\{Aq^5\} +\\
+ [4][a-1][a-2][a-3][a-4]\{Aq^2\}\{Aq^3\}\{Aq^5\}-[a-2][a-3][a-4][a-5]\{Aq^2\}\{Aq^3\}\{Aq^4\}\Big)
  \frac{\{Aq\}}{\{Aq^5\}\{Aq^4\}\{Aq^3\}\{A/q\}}
\!\!\end{array}   }}
\nn \\ \nn \\
3\leq i \leq a:& \alpha^{^{\tiny\boxed{210}}}_{i0} = \alpha_{i0}
\cdot \boxed{\boxed{\frac{[a+1]}{[a-2]}}\cdot \frac{\{Aq^2\}}{\{A/q\}}} \nn \\
& \alpha^{^{\tiny\boxed{210}}}_{01} = \alpha_{01}\cdot \boxed{\boxed{[4]}\cdot\frac{\{A\}}{\{Aq^3\}}}
\nn
\ee
\be
 & \alpha^{^{\tiny\boxed{210}}}_{11} = \alpha_{11}\cdot
\boxed{\frac{\{Aq\}}{\{Aq^3\}}\cdot \boxed{\frac{[4][3][a-1]}{[a]}
\cdot\left(1-\frac{[2][a-3]}{[3][a-1]}\frac{\{Aq^{3}\}}{\{Aq^4\}}\right)}}
\nn \\
& \!\!\!\!\!\!\!\!\!\!\!\!\!\!\!\! \!\!\!\!\!\!\!\!\!\!\!\!\!\!\!\! \!\!\!\!\!\!\!\!\!\!\!\!\!\!\!\!
 \!\!\!\!\!\!\!\!\!\!\!\!\!\!\!\!\!\!\!\!  \alpha^{^{\tiny\boxed{210}}}_{21} = \alpha_{21}\cdot
 \boxed{\frac{\{Aq^2\}}{\{Aq^3\}}\cdot \boxed{
\frac{ [4]\Big([3][a][a-1][a-2]\{Aq^4\}\{Aq^5\}
-[3][a-1][a-2][a-3]\{Aq^3\}\{Aq^5\}+[a-2][a-3][a-4]\{Aq^3\}\{Aq^4\}\Big) }{[a][a-1][a-2]
\cdot\{Aq^4\}\{Aq^5\}}     }}
\nn \\
3\leq i \leq a:& \alpha^{^{\tiny\boxed{210}}}_{i1} = \alpha_{i1}\cdot
\boxed{\boxed{\frac{[4][a+1]}{[a-2]}} }
\nn
\ee
\be
& \ \ \ \ \beta^{\,^{\tiny\boxed{210}}}_{11} =
\frac{\{Aq\}}{\{Aq^3\}}\cdot \alpha_{11}\cdot\frac{\{Aq^{2}\}\{A/q^2\}}{\{Aq\}\{A/q\}}
\cdot\boxed{\frac{\{Aq^2\}}{\{Aq^4\}}\cdot \boxed{\frac{[4][3][a+1]}{[2][a]}\cdot\frac{[1]}{[2]} }}
\nn\\
& \ \ \ \ \beta^{\,^{\tiny\boxed{210}}}_{21} =
\frac{\{Aq\}}{\{Aq^3\}}\cdot \alpha_{11}\cdot\frac{\{Aq^{3}\}\{A/q^2\}}{\{Aq^2\}\{A/q\}}
\cdot\boxed{\frac{\{Aq^3\}}{\{Aq^4\}}\cdot \boxed{\frac{[4][3][a+1]}{[2][a]} \cdot\frac{[2]}{[3]}
\cdot\left([2] - \frac{[a-3]}{[a-1]}\cdot\frac{\{Aq^4\}}{\{Aq^5\}} \right) }}\nn\\
3\leq i \leq a:& \ \ \ \ \beta^{\,^{\tiny\boxed{210}}}_{i1} =
\frac{\{Aq\}}{\{Aq^3\}}\cdot \alpha_{i1}\cdot\frac{\{Aq^{i+1}\}\{A/q^2\}}{\{Aq^i\}\{A/q\}}
\cdot\boxed{\boxed{\frac{[4][3][a+1]}{[2][a-2]}\cdot\frac{[i]}{[i+1]}}}
\nn
\ee
\be
& \ \ \ \ \gamma^{\,^{\tiny\boxed{210}}}_{21} =
\frac{\{A\}}{\{Aq^4\}}\cdot \alpha_{i1}\cdot\frac{\{Aq^{i+2}\}\{A/q^2\}}{\{Aq^i\}\{A\}}
\cdot\boxed{\frac{\{Aq^4\} }{\{Aq^5\} }\cdot \boxed{\frac{[4][a+1]}{[3][a-1]} }} \nn \\
3\leq i \leq a:& \ \ \ \ \gamma^{\,^{\tiny\boxed{210}}}_{i1} =
\frac{\{A\}}{\{Aq^4\}}\cdot \alpha_{i1}\cdot\frac{\{Aq^{i+2}\}\{A/q^2\}}{\{Aq^i\}\{A\}}
\cdot\boxed{  \boxed{\frac{[4][a+1]}{[a-2]}\cdot\frac{[i-1]}{[i+1]}}} \nn \\
3\leq i \leq a:& \ \ \ \ \delta^{\,^{\tiny\boxed{210}}}_{i1} =
\frac{\{Aq\}\{A\}}{\{Aq^4\}\{Aq^3\}}\cdot \alpha_{i1}\cdot\frac{\{Aq^{i+3}\}\{A/q^2\}}{\{Aq^i\}\{Aq\}}
\cdot\boxed{\boxed{\frac{[a+1]}{[a-2]}\cdot\frac{[i-2]}{[i+1]}}}
\label{tbox2box}
\ee

\noindent
and for four boxes --
\be
{\cal F}^{(m)}_{_{\!\!\!\!\!\!\!\!\!\!\!\!\tiny{\boxed{3\ 2\ 1\ 0}
\over{\!\!\!\!\boxed{a\ \ldots \ 3\ 2\ 1\ 0 \ -1}}}}}
\!\!\!\!\!\!
\stackrel{?}{=}\
\frac{q^{6}A^4}{\{Aq^3\}\{Aq^2\}\{Aq\}\{A\}}\cdot\left(
\alpha_\emptyset  +
\sum_{i=0}^a\sum_{j=0}^1 \alpha_{ij}^{^{\tiny\boxed{3210}}} \cdot \lambda_{ij}^{2m}
+ \sum_{i=1}^a \beta_{i1}^{\,^{\tiny\boxed{3210}}}  \cdot
\big(\underbrace{\lambda_{00}\lambda_{i1}}_{\lambda_{\stackrel{00}{i\ 1}}}\big)^{2m}
+ \right. \nn \\ \left.
+ \sum_{i=2}^a \gamma_{i1}^{\,^{\tiny\boxed{3210}}}  \cdot
\big(\underbrace{\lambda_{10}\lambda_{i1}}_{\lambda_{\stackrel{10}{i\ 1}}}\big)^{2m}
+ \sum_{i=3}^a \delta_{i1}^{\,^{\tiny\boxed{3210}}}  \cdot
\big(\underbrace{\lambda_{20}\lambda_{i1}}_{\lambda_{\stackrel{20}{i\ 1}}}\big)^{2m}
+ \sum_{i=4}^a \epsilon_{i1}^{\,^{\tiny\boxed{3210}}}  \cdot
\big(\underbrace{\lambda_{30}\lambda_{i1}}_{\lambda_{\stackrel{30}{i\ 1}}}\big)^{2m}
\right)
\label{F3210}
\ee
with

\be
4\leq i \leq a: \ \ \ \ \ \ \ \ \ \boxed{\alpha^{^{\tiny\boxed{3210}}}_{i0} = \alpha_{i0}
\cdot  { {\frac{[a+1]}{[a-3]}}\cdot \frac{\{Aq^3\}}{\{A/q\}}}
\equiv \tilde \alpha^{^{\tiny\boxed{3210}}}_{i0}}
\nn
\ee
\be
\alpha^{^{\tiny\boxed{3210}}}_{00}\ \stackrel{(\ref{dboxcor})}{ =}\ \alpha_{00}
\cdot  { {\left(1+\frac{[4]}{[a+2]}\frac{\{Aq^a\}}{\{Aq^4\}}\right)}}
=   \tilde\alpha^{^{\tiny\boxed{3210}}}_{00}
\cdot  \frac{1}{[a+2][a+1]}\cdot \frac{\{A/q\}\{Aq^a\}}{\{Aq^3\}}\cdot
 \left(\frac{[a-3][a+2]}{\{Aq^a\}}+\frac{[4][a-3]}{\{Aq^4\}}\right)
 \nn \\ \nn\\
 \!\!\!\!\!\!\!\!\!\!\!\!\!\!\!\!\!\!\!\!\!\!\!\!\!\!\!\!\!\!\!\!\!
 \alpha^{^{\tiny\boxed{3210}}}_{10} =  \tilde\alpha^{^{\tiny\boxed{3210}}}_{10}
\cdot \frac{\{A\} \{Aq^a\}}{\{Aq^3\}}
\cdot \left(\frac{[a-3]}{[a]\,\{Aq^a\}} + \frac{[5][3][a-2][a-3]}{[a][a+1][a+2]\,\{Aq^4\}} -
\frac{[4][3][a-3][a-4]}{[a][a+1][a+2]\,\{Aq^5\}}    \right)
  \nn \\ \nn\\
\!\!\!\!\!\!\!\!\!\!\!\!\!\!\!\!\!\!\!\!\!\!\!\!\!\!\!\!\!\!\!\!
\alpha_{20}^{\,^{\tiny\boxed{3210}}}  = \tilde\alpha_{20}^{\,^{\tiny\boxed{3210}}} \cdot
\frac{\{Aq\}}{[a+2][a+1][a][a-1]}\cdot\left(
\frac{[5][4]}{[2]}\frac{[a][a-1][a-2][a-3]}{\{Aq^3\}}-[5][4]\,\frac{[a-1][a-2][a-3][a-4]}{\{Aq^4\}}+
\right.\nn\\ \left.
+ [5][3]\,\frac{[a-2][a-3][a-4][a-5]}{\{Aq^5\}} -[4]\,\frac{[a-3][a-4][a-5][a-6]}{\{Aq^6\}}
\right)
\nn \\ \nn \\
\!\!\!\!\!\!\!\!\!\!\!\!\!\!\!\!\!\!\!\!\!\!\!\!\!\!
\alpha_{30}^{\,^{\tiny\boxed{3210}}}  = \tilde\alpha_{30}^{\,^{\tiny\boxed{3210}}} \cdot
\frac{\{Aq^2\}}{[a+2][a+1][a][a-1][a-2]} \cdot\left([5]\,\frac{ [a+1][a][a-1][a-2][a-3]}{\{Aq^3\}}
- \frac{[5][4]}{[2]}\,\frac{ [a ][a-1][a-2][a-3][a-4]}{\{Aq^4\}}
+\right. \nn \\ \left.  \!\!\!\!\!\!\!\!\!\!\!\!\!
+\frac{[5][4]}{[2]}\,\frac{[a-1][a-2][a-3][a-4][a-5]}{\{Aq^5\}}
- [5]\,\frac{[a-2][a-3][a-4][a-5][a-6]}{\{Aq^6\}}
+  \frac{[a-3][a-4][a-5][a-6][a-7]}{\{Aq^7\}}
\right)
\nn
\ee
\be
4\leq i \leq a:& \boxed{\alpha^{^{\tiny\boxed{3210}}}_{i1} = \alpha_{i1}\cdot
{{\frac{[5][a+1]}{[a-3]}} }
= \tilde \alpha^{^{\tiny\boxed{3210}}}_{i1}  }
\nn \\ \nn \\
&\alpha_{01}^{\,^{\tiny\boxed{3210}}}  = \tilde\alpha_{01}^{\,^{\tiny\boxed{3210}}} \cdot
\frac{[a-3]}{[a+1]}\frac{\{A \}}{\{Aq^4\}}
\nn \\
&\alpha_{11}^{\,^{\tiny\boxed{3210}}}  = \tilde\alpha_{11}^{\,^{\tiny\boxed{3210}}} \cdot
\frac{[a-3]}{[a+1][a]}\cdot\{Aq\}\left(\frac{[4][a-2]}{\{Aq^4\}}-\frac{[3][a-4]}{\{Aq^5\}}\right)
\nn
\ee
\be
\!\!\!\!\!\!\!\!\!\!\!\!\!\!\!\!\!\!\!\!\!\!\!\!\!\!\!\!\!\!\!
\alpha_{21}^{\,^{\tiny\boxed{3210}}}  = \tilde\alpha_{21}^{\,^{\tiny\boxed{3210}}} \cdot
\frac{\{Aq^2\}}{[a+1][a][a-1]}\cdot\left(
\frac{[4][3]}{[2]}\frac{[a-1][a-2][a-3]}{\{Aq^4\}} - [4][2]\,\frac{[a-2][a-3][a-4]}{\{Aq^5\}}
+ [3]\,\frac{[a_1-3][a_1-4][a_1-5]}{\{Aq^6\}}
\right)
%\frac{[a-3]}{[a][a-1]}\cdot\{Aq^3\}\cdot\left(\frac{[3][a-2]}{\{Aq^5\}}-\frac{[2][a-4]}{\{Aq^6\}}\right)
\nn \\ \nn \\
\!\!\!\!\!\!\!\!\!\!\!\!\!\!\!\!\!\!\!\!\!\!\!\!\!\!\!\!\!\!
\alpha_{31}^{\,^{\tiny\boxed{3210}}}  = \tilde\alpha_{31}^{\,^{\tiny\boxed{3210}}} \cdot
\frac{\{Aq^3\}}{[a+1][a][a-1][a-2] }\cdot \left([4]\cdot \frac{[a][a-1][a-2][a-3]}{\{Aq^4\}}
- \frac{[4][3]}{[2]}\cdot\frac{[a-1][a-2][a-3][a-4]}{\{Aq^5\}}+
\right. \nn \\ \left. +
[4]\cdot\frac{[a-2][a-3][a-4][a-5]}{\{Aq^6\}}
- \frac{[a-3][a-4][a-5][a-6]}{\{Aq^7\}}\right)
\nn
\ee
\be
4\leq i \leq a:& \ \ \ \ \boxed{\beta^{\,^{\tiny\boxed{3210}}}_{i1} =
\frac{\{Aq\}}{\{Aq^4\}}\cdot \alpha_{i1}\cdot\frac{\{Aq^{i+1}\}\{A/q^2\}}{\{Aq^i\}\{A/q\}}
\cdot{{\frac{[5][4][a+1]}{[2][a-3]}\cdot\frac{[i]}{[i+1]}}}
\equiv \tilde \beta^{\,^{\tiny\boxed{3210}}}_{i1}}
\nn\\ \nn \\
&\beta_{11}^{\,^{\tiny\boxed{3210}}}  = \tilde\beta_{11}^{\,^{\tiny\boxed{3210}}} \cdot
\frac{[a-3]}{[a]}\frac{\{Aq^2\}}{\{Aq^5\}}
\nn \\ \nn \\
&\beta_{21}^{\,^{\tiny\boxed{3210}}}  = \tilde\beta_{21}^{\,^{\tiny\boxed{3210}}} \cdot
\frac{[a-3]}{[a][a-1]}\cdot\{Aq^3\}\cdot\left(\frac{[3][a-2]}{\{Aq^5\}}-\frac{[2][a-4]}{\{Aq^6\}}\right)
\nn
\ee
\be
\!\!\!\!\!\!
\beta_{31}^{\,^{\tiny\boxed{3210}}}  = \tilde\beta_{31}^{\,^{\tiny\boxed{3210}}} \cdot
\frac{\{Aq^4\}}{[a][a-1][a-2]}\cdot\left([3]\cdot\frac{ [a-1][a-2][a-3]}{\{Aq^5\}}
-[3]\cdot \frac{[a-2][a-3][a-4]}{\{Aq^6\}}+\frac{[a-3][a-4][a-5]}{\{Aq^7\}}\right)
\nn
\ee
\be
4\leq i \leq a:& \ \ \ \ \boxed{\gamma^{\,^{\tiny\boxed{3210}}}_{i1} =
\frac{\{Aq^3\}\{A\}}{\{Aq^5\}\{Aq^4\}}\cdot \alpha_{i1}\cdot\frac{\{Aq^{i+2}\}\{A/q^2\}}{\{Aq^i\}\{A\}}
\cdot{{\frac{[5][4][a+1]}{[2][a-3]}\cdot\frac{[i-1]}{[i+1]}}}
\equiv \tilde \gamma^{\,^{\tiny\boxed{3210}}}_{i1}}
\nn \\ \nn \\
&\gamma_{21}^{\,^{\tiny\boxed{3210}}}  = \tilde\gamma_{21}^{\,^{\tiny\boxed{3210}}} \cdot
\frac{[a-3]}{[a-1]}\frac{\{Aq^4\}}{\{Aq^6\}}
\nn \\ \nn \\
&\gamma_{31}^{\,^{\tiny\boxed{3210}}}  = \tilde\gamma_{31}^{\,^{\tiny\boxed{3210}}} \cdot
\frac{[a-3]}{[a-1][a-2]}\cdot\{Aq^5\}\cdot\left(\frac{[2][a-2]}{\{Aq^6\}}-\frac{[a-4]}{\{Aq^7\}}\right)
\nn \\ \nn \\
4\leq i \leq a:& \ \ \ \ \boxed{\delta^{\,^{\tiny\boxed{3210}}}_{i1} =
\frac{\{Aq\}\{A\}}{\{Aq^6\}\{Aq^4\}}\cdot \alpha_{i1}\cdot\frac{\{Aq^{i+3}\}\{A/q^2\}}{\{Aq^i\}\{Aq\}}
\cdot{{\frac{[5][a+1]}{[a-3]}\cdot\frac{[i-2]}{[i+1]}}}
\equiv \tilde \delta^{\,^{\tiny\boxed{3210}}}_{i1}  }
\nn\\ \nn \\
&\delta_{31}^{\,^{\tiny\boxed{3210}}}  = \tilde\delta_{31}^{\,^{\tiny\boxed{3210}}} \cdot
\frac{[a-3]}{[a-2]}\frac{\{Aq^6\}}{\{Aq^7\}}
\nn \\ \nn \\
4\leq i \leq a:& \ \ \ \ \boxed{\epsilon^{\,^{\tiny\boxed{3210}}}_{i1} =
\frac{\{Aq^2\}\{Aq\}\{A\}}{\{Aq^6\}\{Aq^5\}\{Aq^4\}}\cdot \alpha_{i1}
\cdot\frac{\{Aq^{i+4}\}\{A/q^2\}}{\{Aq^i\}\{Aq^2\}}
\cdot{{\frac{[a+1]}{[a-3]}\cdot\frac{[i-3]}{[i+1]}}}
\equiv  \tilde \epsilon^{\,^{\tiny\boxed{3210}}}_{i1}}
\label{qbox2box}
\ee

\noindent
These formulas are already sufficient to handle the case of $R=[55]$,
where only  $a_2<a<5$, i.e. $a_2\leq 3$ are needed.

\subsection*{Generic two-floor case with $b_1=1$}

Even more important, the {\it structure} of formulas is now getting clear
and this opens the way to higher $r>5$.
First, we can note the for $i\geq 4$ expressions {\it stabilize} and are easily described in full generality
- in (\ref{qbox2box}) they are put in boxes.
Second, one can continue these stable expressions to $i_2<i\leq 3=a_2$
-- in (\ref{qbox2box}) continuation is denoted by tilde --
and separate the interpolating/correction factors  which then also have a pronounced structure.

The last step towards  general formula for the two-floor pyramids is to provide expressions
which unify   (\ref{firstfloorconj}), (\ref{dboxmin}), (\ref{F3210}),  (\ref{qbox2box})
with different values of $a_2$ into a single formula.
We keep restriction to the case of $b_1=0$, which is relevant for representations $R=[rr]$ and
its transposed $R=[2^r]$,
but inclusion of arbitrary $b_2<b_1$ is also straightforward.
As in the last formula (\ref{qbox2box}) above,
we express it in terms of the single-floor functions (\ref{firstfloorconj}):
\be
{\cal F}^{(m)}_{\ \ \ \ \ \ \ \ \ _{\!\!\!\!\!\!\!\!\!\!\!\!\!\!\!\!\!\!\!\!\!\!\!\!\!\!\!\!\!\!\!\!\!\!\!
\tiny{\boxed{a_2\ \ldots\ 0}
\over{\boxed{a_1\ \ \ \ \ldots \  \ \ \ 0 \ -1}}}}}
\!\!\!\!\!
\stackrel{?}{=}\
\frac{q^{\frac{a_2(a_2+1)}{2}}A^{a_2+1}}{\prod_{j=0}^{a_2} \{Aq^j\}}
 \cdot\left(
\alpha_\emptyset  \ + \
\sum_{i=0}^{a_1}\sum_{j=0}^1 \alpha_{ij}^{^{\tiny\boxed{a_2\ldots 0}}} \cdot \lambda_{ij}^{2m}
%+ \sum_{i=0}^a \alpha_{i1}^{^{\tiny\boxed{a_2\ldots 0}}} \cdot \lambda_{i1}^{2m}
\ + \  \sum_{i_2=0}^{a_2}\sum_{i_1=1}^{a_1} B_{_{i_2i_1}}^{\,^{\tiny\boxed{a_2\ldots 0}}}  \cdot
\big(\underbrace{\lambda_{i_20}\lambda_{i_11}}_{\lambda_{\stackrel{i_2\ 0}{i_1\ \ 1}}}\big)^{2m}
\right)
\label{2floorF}
\ee
Note the unifying notation: the previous
$\beta_{i_1\,1}=B_{0\,i_1}, \
\gamma_{i_1\,1}=B_{1\,i_1}, \ \delta_{i_1\,1}=B_{2\,i_1},\ \epsilon_{i_1\,1}=B_{\,i_1}, \ldots$
The simplest are the general formulas for the "stable" coefficients with $a_2< i_1 \leq a_1$,
when the rules (\ref{boxrule2}) are directly applicable:
\be
a_2< i_1 \leq a_1:
%\ \ \ \ \ \ \ \
&\boxed{\alpha^{^{\tiny\boxed{a_2\ldots 0}}}_{i\,0} =
\alpha_{i\,0}
\cdot {{\frac{[a_1+1]}{[a_1-a_2]}}\cdot \frac{\{Aq^{a_2}\}}{\{A/q\}}}
\equiv   \tilde \alpha^{^{\tiny\boxed{a_2\ldots 0}}}_{i\,0} }
\nn \\ \nn \\
a_2< i \leq a:
%\ \ \ \ \ \ \ \ \ \
&\boxed{\alpha^{^{\tiny\boxed{a_2\ldots 0}}}_{i\,1}
=\alpha_{i\,1}\cdot
{{[a_2+2]\cdot\frac{[a_1+1]}{[a_1-a_2]}} } \equiv  \tilde \alpha^{^{\tiny\boxed{a_2\ldots 0}}}_{i\,1}}
\label{2flooralpha}
\ee
\be
\!\!\!\!\!\!\!\!
\boxed{  B^{^{\,\tiny\boxed{a_2\ldots 0}}}_{i\,1} =
  \frac{\{Aq^{2i_2+1}\}\prod_{j=0}^{i_2-1}\{Aq^j\}}{\prod_{j=0}^{i_2}\{Aq^{a_2+j+1}\}}
\cdot   \alpha_{i1}
\cdot \frac{\{Aq^{i_1+i_2+1}\}\{A/q^2\}_{\phantom{5_5}}\!\!\!}
{\{Aq^{i_1}\}\{Aq^{i_2-1}\}^{\phantom{5^5}}\!\!\!}
 \cdot{{\frac{[a_2+2]!}{[a_2-i_2]![i_2+2]!}\cdot
\frac{[a_1+1] }{[a_1-a_2]}\cdot\frac{[i_1-i_2]}{[i_1+1]} }}
\equiv \tilde B^{^{\,\tiny\boxed{a_2\ldots 0}}}_{i\,1}  }
\nn
\ee
Note that in fact $\alpha^{^{\tiny\boxed{a_2\ldots 0}}}_{i0}$ is also given by the last
formula, where one should just put $i_2+1=0$.
This means that assuming $\lambda_{-1,0}=\lambda_\emptyset = 1$,
one can rewrite (\ref{2floorF}) as
\be
{\cal F}^{(m)}_{\ \ \ \ \ \ \ \ \ _{\!\!\!\!\!\!\!\!\!\!\!\!\!\!\!\!\!\!\!\!\!\!\!\!\!\!\!\!\!\!\!\!\!\!\!
\tiny{\boxed{a_2\ \ldots\ 0}
\over{\boxed{a_1\ \ \ \ \ldots \  \ \ \ 0 \ -1}}}}}
\!\!\!\!\!
\stackrel{?}{=}\
\frac{q^{\frac{a_2(a_2+1)}{2}}A^{a_2+1}}{\prod_{j=0}^{a_2} \{Aq^j\}}
 \cdot\left(
\alpha_\emptyset  \ + \
\sum_{i=0}^{a_1}  \alpha_{i0}^{^{\tiny\boxed{a_2\ldots 0}}} \cdot \lambda_{i0}^{2m}
%+ \sum_{i=0}^a \alpha_{i1}^{^{\tiny\boxed{a_2\ldots 0}}} \cdot \lambda_{i1}^{2m}
+ \sum_{i_2=-1}^{a_2}\sum_{i_1=1}^{a_1} B_{_{i_2i_1}}^{\,^{\tiny\boxed{a_2\ldots 0}}}  \cdot
\big(\underbrace{\lambda_{i_20}\lambda_{i_11}}_{\lambda_{\stackrel{i_2\ 0}{i_1\ \ 1}}}\big)^{2m}
\right)
\label{2floorFmod}
\ee
Eq.(\ref{2flooralpha}) should now be supplemented by the list of correction factors
at $i_1\leq a_2$, like
\be
&\alpha^{^{\tiny\boxed{a_2\ldots 0}}}_{00}\ \stackrel{(\ref{dboxcor})}{ =}\  \alpha_{00}
\cdot {{\left(1+\frac{[a_2+1]}{[a_1+2]}\frac{\{Aq^{a_1}\}}{\{Aq^{a_2+1}\}}\right)}}
\nn \\ \nn \\
&\!\!\!\!\!\!\!\!\!\!\!\!\!\!\!\!\!\!\!\!\!\!\!\!\!\!\!\!\!\!\!\!\!\!\!\!\!\!\!\! \!\!\!\!
 \alpha^{^{\tiny\boxed{a_2\ldots 0}}}_{10} \!= \alpha_{10}
\cdot {\frac{\{A\}}{\{A/q\}}\cdot {
\left(\frac{[a_1+1]}{[a_1]} +
 \frac{[a_2+2][a_2]\cdot [a_{12}+1]}{[a_1+2][a_1] }\frac{\{Aq^{a_1}\}}{\{Aq^{a_2+1}\}}
-  \frac{[a_2+1][a_2]\cdot [a_{12}-1]}{[a_1+2][a_1] }\frac{\{Aq^{a_1}\}}{\{Aq^{a_2+2}\}}
\right)}}
\nn \\ \nn \\
&\alpha^{^{\tiny\boxed{a_2\ldots 0}}}_{01} = \alpha_{01}\cdot {{[a_2+2]}
\cdot\frac{\{A\}}{\{Aq^{a_2+1}\}}}
\nn \\
& \ldots
\ee
If expressed through continuation of the stable expressions (\ref{2flooralpha}),
they automatically turn into unities for $a_2<i_1\leq a_1$:
\be
 \alpha^{^{\tiny\boxed{a_2\ldots 0}}}_{00}\ \stackrel{(\ref{dboxcor})}{ =}\
 \tilde\alpha^{^{\tiny\boxed{a_2\ldots 0}}}_{00}
\cdot  \frac{1}{[a_1+2][a_1+1]}\cdot \frac{\{A/q\}\{Aq^{a_1}\}}{\{Aq^{a_2}\}}\cdot
 \left(\frac{[a_{12}]\cdot [a_1+2]}{\{Aq^{a_1}\}}+\frac{[a_2+1]\cdot [a_{12}]}{\{Aq^{a_2+1}\}}\right)
 \nn \\ \nn \\
 \!\!\!\!\!\!\!\!\!\!\!\!\!\!\!\!\!\!\!\!\!\!\!\!\!\!\!\!\!\!\!\!\!
 \alpha^{^{\tiny\boxed{a_2\ldots 0}}}_{10} =  \tilde\alpha^{^{\tiny\boxed{a_2\ldots 0}}}_{10}
\cdot \frac{\{A\} \{Aq^{a_1}\}}{\{Aq^{a_2}\}}
\cdot \left(\frac{[a_{12}]}{[a_1]\,\{Aq^{a_1}\}}
+ \frac{[a_2+2][a_2]\cdot[a_{12}+1][a_{12}]}{[a_1+2][a_1+1][a_1]\,\{Aq^{a_2+1}\}} -
\frac{[a_2+1][a_2]\cdot [a_{12}][a_{12}-1]}{[a_1+2][a_1+1][a_1]\,\{Aq^{a_2+2}\}}    \right)
  \nn \\ \nn \\
\!\!\!\!\!\!\!\!\!\!\!\!\!\!\!\!\!\!\!\!\!\!\!\!\!\!\!\!\!\!\!\!
\alpha_{20}^{\,^{\tiny\boxed{a_2\ldots 0}}}  = \tilde\alpha_{20}^{\,^{\tiny\boxed{a_2\ldots 0}}} \cdot
\frac{\{Aq\}}{[a_1+2][a_1+1][a_1][a_1-1]}\cdot\left(
\frac{[a_2+2][a_2+1][a_2]}{[3][2]}\cdot\frac{[a_{12}+3][a_{12}+2][a_{12}+1][a_{12}]}{\{Aq^{a_2}\}}-
\right. \nn \\
-\frac{[a_2+2][a_2+1][a_2-1]}{[2]}\cdot\frac{[a_{12}+2][a_{12}+1][a_{12}][a_{12}-1]}{\{Aq^{a_2+1}\}}
+ \frac{[a_2+2][a_2][a_2-1]}{[2]}\cdot\frac{[a_{12}+1][a_{12}][a_{12}-1][a_{12}-2]}{\{Aq^{a_2+2}\}}-
\nn\\ \left.
-\frac{[a_2+1][a_2][a_2-1]}{[3][2]}\cdot\frac{[a_{12}][a_{12}-1][a_{12}-2][a_{12}-3]}{\{Aq^{a_2+3}\}}
\right)
\nn
\ee
{\footnotesize
\be
%\!\!\!\!\!\!\!\!\!\!\!\!\!\!\!\!\!\!\!\!\!\!\!\!\!\!\!\!\!\!\!\!\!
\alpha_{30}^{\,^{\tiny\boxed{a_2\ldots 0}}}  = \tilde\alpha_{30}^{\,^{\tiny\boxed{a_2\ldots 0}}} \cdot
\frac{\{Aq^2\}}{[a_1+2][a_1+1][a_1][a_1-1][a_1-2]}
\cdot\left(\frac{[a_2+2][a_2+1][a_2][a_2-1]}{[4][3][2]}
\cdot\frac{ [a_{12}+4][a_{12}+3][a_{12}+2][a_{12}+1][a_{12}]}{\{Aq^{a_2}\}}
-\right. \nn \\ \left.
 \!\!\!\!\!\!\!\!\!\!\!\!\!\!\!\!\!\!\!\!\!\!\!\!\!\!\!
 %\!\!\!\!\!\!\!\!\!\!\!\!
%\!\!\!\!\!\!\!\!\!\!\!\!\!
- \frac{[a_2+2][a_2+1][a_2][a_2-2]}{[3][2]}
\cdot\frac{ [a_{12}+3][a_{12}+2][a_{12}+1][a_{12}][a_{12}-1]}{\{Aq^{a_2+1}\}}
+\frac{[a_2+2][a_2+1][a_2-1][a_2-2]}{[2]^2}
\cdot \frac{[a_{12}+2][a_{12}+1][a_{12}][a_{12}-1][a_{12}-2]}{\{Aq^{a_2+2}\}}
- \right. \nn \\ \left.  \!\!\!\!\!\!\!\!\!\!\!\!\!\!\!\!\!\!\!\!\!\!\!\!\!\!
- \frac{[a_2+2][a_2][a_2-1][a_2-1]}{[3][2]}
\cdot \frac{[a_{12}+1][a_{12}][a_{12}-1][a_{12}-2][a_{12}-3]}{\{Aq^{a_2+3}\}}
+  \frac{[a_2+1][a_2][a_2-1][a_2-2]}{[4][3][2]}\cdot
\frac{[a_{12}][a_{12}-1][a_{12}-2][a_{12}-3][a_{12}-4]}{\{Aq^{a_2+4}\}}
\right)
\nn\\ \nn \\
\ldots \nn
\ee
}
\be
\alpha_{01}^{\,^{\tiny\boxed{a_2\ldots 0}}}  = \tilde\alpha_{01}^{\,^{\tiny\boxed{a_2\ldots 0}}} \cdot
\frac{[a_{12}]}{[a_1+1]}\frac{\{A \}}{\{Aq^{a_2+1}\}}
\nn \\ \nn \\
\alpha_{11}^{\,^{\tiny\boxed{a_2\ldots 0}}}  = \tilde\alpha_{11}^{\,^{\tiny\boxed{a_2\ldots 0}}} \cdot
\frac{[a_{12}]}{[a_1+1][a_1]}\cdot\{Aq\}\cdot
\left(\frac{[a_2+1]\cdot[a_{12}+1]}{\{Aq^{a_2+1}\}}-\frac{[a_2]\cdot[a_{12}-1]}{\{Aq^{a_2+2}\}}\right)
\nn \\ \nn \\
%\!\!\!\!\!\!\!\!\!\!\!\!\!\!\!\!\!\!\!\!\!\!\!\!\!\!\!\!\!\!\!
\alpha_{21}^{\,^{\tiny\boxed{a_2\ldots 0}}}  = \tilde\alpha_{21}^{\,^{\tiny\boxed{a_2\ldots 0}}} \cdot
\frac{\{Aq^2\}}{[a_1+1][a_1][a_1-1]}\cdot\left(
\frac{[a_2+1][a_2]}{[2]}\cdot\frac{[a_{12}+2][a_{12}+1][a_{12}]}{\{Aq^{a_2+1}\}} -
\right. \nn \\ \left.
- [a_2+1][a_2-1]\cdot\frac{[a_{12}+1][a_{12}][a_{12}-1]}{\{Aq^{a_2+2}\}}
+ \frac{[a_2][a_2-1]}{[2]}\cdot\frac{[a_{12}][a_{12}-1][a_{12}-2]}{\{Aq^{a_2+3}\}}
\right)
 \nn \\ \nn \\
\alpha_{31}^{\,^{\tiny\boxed{3210}}}  = \tilde\alpha_{31}^{\,^{\tiny\boxed{3210}}} \cdot
\frac{\{Aq^3\}}{[a+1][a_1][a_1-1][a_1-2] }
\cdot \left(\frac{[a_2+1][a_2][a_2-1]}{[3][2]}\cdot \frac{[a_{12}+3][a_{12}+2][a_{12}+1][a_{12}][}{\{Aq^{a_2+1}\}} -
\right. \nn \\ \left.
%\!\!\!\!\!\!\!\!\!\!\!\!\!\!\!\!\!\!\!\!\!\!
- \ \frac{[a_2+1][a_2][a_2-2]}{[2]}\cdot\frac{[a_{12}+2][a_{12}+1][a_{12}][a_{12}-1]}{\{Aq^{a_2+2}\}}
+\ \frac{[a_2+1][a_2-1][a_2-2]}{[2]}\cdot\frac{[a_{12}+1][a_{12}][a_{12}-1][a_{12}-2]}{\{Aq^{a_2+3}\}}-
\right. \nn \\ \left.
- \frac{[a_2][a_2-1][a_2-2]}{[3][2]}\frac{[a_{12}][a_{12}-1][a_{12}-2][a_{12}-3]}{\{Aq^{a2+4}\}}\right)
\nn\\ \nn \\
\ldots \nn
\ee
\be
\beta_{11}^{\,^{\tiny\boxed{a_2\ldots 0}}}  = \tilde\beta_{11}^{\,^{\tiny\boxed{a_2\ldots 0}}} \cdot
\frac{[a_{12}]}{[a_1]}\frac{\{Aq^2\}}{\{Aq^{a_2+2}\}} \nn \\ \nn \\
\beta_{21}^{\,^{\tiny\boxed{a_2\ldots 0}}}  = \tilde\beta_{21}^{\,^{\tiny\boxed{a_2\ldots 0}}} \cdot
\frac{[a_{12}]}{[a_1][a_1-1]}\cdot\{Aq^3\}\cdot
\left(\frac{[a_2]\cdot[a_{12}+1]}{\{Aq^{a_2+2}\}}-\frac{[a_2-1]\cdot[a_{12}-1]}{\{Aq^{a_2+3}\}}\right)
\nn \\ \nn\\
%\!\!\!\!\!\!\!\!\!\!\!\!\!\!\!\!\!\!\!\!\!\!
\beta_{31}^{\,^{\tiny\boxed{3210}}}  = \tilde\beta_{31}^{\,^{\tiny\boxed{3210}}} \cdot
\frac{\{Aq^4\}}{[a_1][a_1-1][a_1-2]}
\cdot\left(\frac{[a_2][a_2-1]}{[2]}\cdot\frac{ [a_{12}+2][a_{12}+1][a_{12}]}{\{Aq^{a_2+2}\}} -
\ \ \ \ \ \ \ \ \ \ \
\right.\nn\\ \left.
 -[a_2][a_2-2]\cdot \frac{[a_{12}+1][a_{12}][a_{12}-1]}{\{Aq^{a_2+3}\}}
+\frac{[a_2-1][a_2-2]}{[2]}\frac{[a_{12}][a_{12}-1][a_{12}-2]}{\{Aq^{a_2+4}\}}\right)
\nn\\ \nn \\
\ldots \nn
\ee
\be
\gamma_{21}^{\,^{\tiny\boxed{a_2\ldots 0}}}  = \tilde\gamma_{21}^{\,^{\tiny\boxed{a_2\ldots 0}}} \cdot
\frac{[a_{12}]}{[a_1-1]}\frac{\{Aq^4\}}{\{Aq^{a_2+3}\}} \nn \\
\gamma_{31}^{\,^{\tiny\boxed{a_2\ldots 0}}}  = \tilde\gamma_{31}^{\,^{\tiny\boxed{a_2\ldots 0}}} \cdot
\frac{[a_{12}]}{[a_1-1][a_2-2]}\cdot\{Aq^5\}\cdot
\left(\frac{[a_2-1]\cdot [a_{12}+1]}{\{Aq^{a_2+3}\}}-\frac{[a_2-2]\cdot[a_{12}-1]}{\{Aq^{a_2+4}\}}\right)
\nn \\ \nn \\
\ldots \nn
\ee
\be
\delta_{31}^{\,^{\tiny\boxed{a_2\ldots 0}}}  = \tilde\delta_{31}^{\,^{\tiny\boxed{a_2\ldots 0}}} \cdot
\frac{[a_1-a_2]}{[a_1-2]}\frac{\{Aq^6\}}{\{Aq^{a_2+4}\}}
\nn \\ \nn \\
\ldots
\label{gencorr}
\ee
We introduced the abbreviated notation $a_{12}=a_1-a_2$ to simplify the formulas,
at least a little.
From this list it is clear that the combinatorial coefficients in the previous
formulas  become explicit functions of $a_2$.
%The coefficients in boxes still need to be converted into this form --
%given are their values at $a_2=3$, borrowed from (\ref{qbox2box})).
Continuing the list of correction coefficients, as well as  generalizations beyond two floors
are now straightforward.

\subsection*{Calculational tricks and proof directions}

Once their structure is understood,
the simplest way to deduce above formulas for ${\cal F}_{\A,\B}^{(m)}$ is to write them in the form
\be
\left(\prod_f \prod_{j=-2b_f}^{2a_f} \{Aq^j\}\right) \cdot {\cal F}_{\A,\B}^{(m)}
= \prod_f \left(\prod_{j=-2b_f}^{-(b_f+1)} \{Aq^j\}\prod_{j=a_f+1}^{2a_f} \{Aq^j\}\right)
\ +\ \sum_{\I,\J}  P_{\I,\J}(A,q)\cdot \lambda_{\I,\J}^{2m}
\label{Fpol}
\ee
with the first term associated with the {\it complement} of the pyramid
and $P_{\I,\J}$ some Laurent polynomials of the same degree
\be
\sum_f \frac{2a_f(2a_f+1)-(a_f+1)(a_f+2)+2b_f(2b_f+1)-(b_f+1)(b_f+2)}{2}
\ee
in $A$ with $q$-dependent coefficients.
These polynomials do not depend on $m$ and at the same time the r.h.s. of (\ref{Fpol}) should
vanish at all zeroes of the pre-factor in the l.h.s. -- because ${\cal F}^{(m)}$ itself should
be a Laurent polynomial in $A$ and $q$ at all $m$, though represented as a sum of rational functions.
This constraint provides an infinite system of linear equations on the large, but finite
set of $q$-dependent coefficients of polynomials $P_{\I,\J}$,
which can be straightforwardly solved.
The conjecture is that it is a consistent system (solution exists, which solves all the vanishing
constraints for infinitely many $m$) and it is unique.
In fact, if one wishes to {\it prove} these formulas for ${\cal F}$ -- the simplest thing
is to prove this linear-algebra theorem.

The key point here is to notice that the set (\ref{eigenvalues})
of eigenvalues $\lambda_{\I,\J}$ is somewhat special:
at each point $A=q^{n}$, which is a zero of denominator, some of $\lambda$'s coincide.
Thus for the numerator of ${\cal F}$ to vanish,
only sums of the coefficients in front of coinciding eigenvalues
should vanish, not the coefficients themselves -- and this allows for a non-trivial solution.
As an elementary illustration, in
\be
{\cal F}_{_{\tiny\boxed{210}}}^{(m)}\ \sim \  \frac{\alpha_\emptyset(A,q)\cdot \lambda_\emptyset^{2m}
\ +\ \alpha_{00}(A,q)\cdot \lambda_{00}^{2m}
\ +\ \alpha_{10}(A,q)\cdot\lambda_{10}^{2m} }{\{Aq^2\}\{Aq\}\{A\}}
\label{F210sample}
\ee
with $\lambda_\emptyset=1$, $\lambda_{00}=-A$ and $\lambda_{10}=q^2A^2\ $
we have at the three poles of denominator:
\be
\begin{array}{ccccccc}
A=1 &   \lambda_{00}^2=\lambda_\emptyset^2 &\Longrightarrow &
\alpha_\emptyset+\alpha_{00}=0 \ \& \ \alpha_{10}=0  &\Longrightarrow & \alpha_{10}\sim \{A\} \\ \\
A=q^{-1} &   \lambda_{10}^2=\lambda_\emptyset^2 &\Longrightarrow &
\alpha_\emptyset+\alpha_{10}=0 \ \& \ \alpha_{00}=0  &\Longrightarrow & \alpha_{00}\sim \{Aq\} \\ \\
A=q^{-2} &   \lambda_{10}^2=\lambda_{00}^2 &\Longrightarrow &
\alpha_{00}+\alpha_{10}=0 \ \& \ \alpha_\emptyset=0  &\Longrightarrow & \alpha_\emptyset\sim \{Aq^2\} \\
\end{array}
\ee
and solving remaining constraints we find that the numerator is
proportional to $\{Aq^2\}-[2]\{Aq\}\cdot \lambda_{00}^{2m} + \{A\}\cdot \lambda_{10}^{2m}$,
while common factor in front of it is fixed by the known coefficient in front of $\lambda_\emptyset^{2m}=1$.
Solvability and uniqueness are guaranteed by the properties of the matrix
$\left(\begin{array}{ccc} 1&-1&0 \\ 1&0&-1 \\0&1&-1 \end{array}\right)$, characterizing relations between
eigenvalues at roots, -- it has corank one.
In the same way one can handle expressions for more complicated pyramids.
In fact, this linear-algebra theorem can be used as an alternative {\it definition} of ${\cal F}$.

\bigskip

However, as often happens, the proof itself does not provide explicit formulas, like those in this text.
Technically, the simplest way to get them is to solve the linear system at particular value
of $q$ (in practice, $q=2$ is enough).
Since most polynomials $P_{\I,\J}$ are actually factorized into products of differentials
$\{Aq^j\}$ and the main problem is to associate a set of exponents $j$ with each pair of
embedded pyramids $\{\I,\J\}\subset\{\A,\B\}$, the $q$-dependence is easily restored.
Non-factorized terms in double boxes in above formulas can after that be recalculated
by keeping $q$ arbitrary -- but at this stage the number of unknown coefficients is already small.

\subsection*{Towards Racah matrices}

Given explicit expressions (\ref{321010}), (\ref{3210110}), (\ref{32101210})
one can deduce exclusive Racah matrices for $R=[4,4]$.
Formulating more general rules like
%(\ref{dboxmin}), (\ref{boxrule2}), (\ref{dboxcor}) etc
(\ref{2flooralpha})-(\ref{gencorr})
one can proceed further to $R=[r,r]$ with either concrete or generic $r$.
For $R=[r^s]$ one needs to consider $s$-floor pyramids, but
in fact, the few first examples are enough to conjecture formulas for
exclusive Racah in generic rectangular representations.
However, it is a separate story, beyond the scope of the present paper.

\section*{Acknowledgements}

This work was performed at the Institute for the Information Transmission Problems
with the  support from the Russian Science Foundation, Grant No.14-50-00150.

\end{document}